\documentclass[11pt]{article}

\usepackage[margin=30mm]{geometry}
\usepackage{amsmath}
\usepackage[pdftex]{graphicx}
\usepackage{subfig}

\usepackage{authblk}

\newcommand{\mC}[1]{\mathcal{#1}}

\title{A numerical study of observational systematic errors in lensing analysis of CMB polarization}

\author[1]{Ryo Nagata}

\author[2]{Toshiya Namikawa}

\affil[1]{Japan Aerospace Exploration Agency (JAXA), Institute of Space and Astronautical Science (ISAS), Sagamihara, Kanagawa  252-5210, Japan 
}

\affil[2]{Department of Applied Mathematics and Theoretical Physics, The University of Cambridge, Wilberforce Rd, Cambridge CB3 0WA, United Kingdom}

\begin{document}

\maketitle

\begin{abstract}%
Impacts of observational systematic errors on the lensing analysis of the cosmic microwave background (CMB) polarization are investigated by numerical simulations. We model errors of gain, angle, and pointing in observation of the CMB polarization and simulate polarization fields modulated by the errors. We discuss the response of systematics-induced $B$-modes to amplitude and spatial scale of the imposed errors and show that the results of the lensing reconstruction and delensing analysis behave according to it. It is observed that error levels expected in the near future lead to no significant degradation in delensing efficiency. 
\end{abstract}

\section{Introduction} \label{introduction}
Detection of the primordial gravitational waves (GWs) which originate from cosmic inflation is expected to provide us with an opportunity to understand a very early stage of the universe. An attempt to extract a signal of the primordial GWs from $B$-modes of the cosmic microwave background (CMB) polarization is one of the most promising ways to this end \cite{Polnarev:1985,Kamionkowski:1996:GW}. 
The tightest bound to-date on the amplitude of the primordial GWs characterized by the tensor-to-scalar ratio, $r$, is $r<0.06$ ($95\%$ C.L.) obtained from the BICEP2/Keck Array \cite{BKX} or $r<0.044$ ($95\%$ C.L.) from the BICEP2/Keck Array and Planck \cite{Planck-NPIPE}. 

In recent observations of the CMB polarization, employment of integrated detector arrays is rapidly increasing measurement sensitivity. Observations in the near future will provide low-noise polarization maps of a few $\mu$K-arcmin \cite{SO2019,PICO2019,S42019,CMB-HD}. It is expected that a major contaminant in observed polarization maps is disturbance due to gravitational lensing effects \cite{Kesden:2002ku,Seljak:2003pn}. 
Gravitational lensing converts a small part of $E$-modes into $B$-modes \cite{Zaldarriaga:1998:LensB}. The lensing $B$-modes have a spectrum like white noise in large scales with an amplitude comparable to the $B$-mode spectrum of the primordial GWs origin with $r=0.01$ at the scale of the recombination bump. 
Delensing by which we remove the contamination from the lensing effects is considered to be an indispensable analysis method in next-generation high-sensitivity experiments \cite{Smith:2010gu,Carron:2018:GW,Millea:2020:MLEphi,CMBS4:r-forecast-rev}. 

To achieve delensing, we need to identify the lensing mass distribution, which is characterized by the quantity called ``lensing potential''. CMB lensing mass maps have been reconstructed by multiple CMB observations such as ACTPol \cite{ACT:phi:2011,ACT16:phi}, BICEP2/Keck Array \cite{BKVIII}, Planck \cite{P13:phi,P15:phi,P18:phi}, POLARBEAR \cite{PB:phi:2014,PB:phi:2019} and SPTpol \cite{SPT:phi:2015,SPT:phi:2019}. Ongoing, near-term, and next-generation CMB observations such as Advanced ACT \cite{AdvACT}, Simons Observatory \cite{SO2019}, SPT-3G \cite{SPT3G}, and CMB Stage-IV \cite{CMBS4:r-forecast-rev} will significantly increase sensitivity to the lensing potential. Those reconstructed mass maps can be used to form a lensing $B$-mode template which is then subtracted from observed $B$-modes \cite{Seljak:2003pn} or is incorporated into a likelihood analysis as a new ``frequency'' map \cite{BKSPT}. In addition to the lensing maps measured internally with CMB data, we can also use external mass tracers which correlate with the CMB lensing signal efficiently, such as the cosmic infrared background (CIB) \cite{Simard:2015,Sherwin:2015,Yu:2017:multi-tracer}, radio and optical galaxies \cite{Namikawa:2015:delens,Manzotti:2018}, galaxy weak lensing \cite{Marian:2007}, and intensity 
maps of high-redshift line emissions \cite{Sigurdson:2005cp,Karkare:2019:delens}. In the last few years, several works have demonstrated delensing for small-scale lensed temperature and polarization anisotropies using real data \cite{SPTpol:2017:delens,Larsen:2016,Carron:2017,PB:2019:delens,DW:2020:delens}. Recently, Ref.~\cite{BKSPT} demonstrated for the first time delensing of large-scale $B$-modes using CIB as a mass tracer and obtained an improved constraint on $r$ compared with that without delensing. 

Multiple works have explored possible systematics in delensing for future CMB experiments. 
For example, delensed CMB anisotropies in small scales have large biases arising from higher-order correlations in the CMB anisotropies \cite{Namikawa:2017:delens} and correlation between the lensing template and the CMB anisotropies to be delensed \cite{Teng:2011xc}. This bias can be mitigated by several proposed methods \cite{PB:2019:delens,Namikawa:2017:delens,Sehgal:2016}. To constrain the primordial GWs, we need only large-scale $B$-modes and can simply mitigate this bias by using only small-scale CMB anisotropies to construct the lensing $B$-mode template \cite{Seljak:2003pn,Namikawa2014a,Lizancos:2020:delens-bias}. 
Also, the lensing $B$-modes are known to be a non-Gaussian field and non-Gaussian covariance of the lensing $B$-mode power spectrum degrades cosmological parameter constraints \cite{BenoitLevy:2012va}.  However, non-Gaussian covariance of the delensed $B$-mode power spectrum, which potentially degrades a constraint on $r$, is negligible in large angular scales \cite{Namikawa:2015:cov}. 
A power spectrum of $B$-modes delensed by using external mass tracers could be biased by uncertainties of the tracers themselves. However, the delensed $B$-modes would not have significant contributions from non-linear growth of the large-scale structure \cite{Namikawa:2018:nldelens} and residual Galactic foreground emissions \cite{ref:cib_delensing_biases}. Further, the uncertainties of the mass tracers are, in general, constrained by power spectra of the mass tracers as well as cross-spectra between the mass tracers and CMB lensing maps \cite{Sherwin:2015}. In actual observations, the presence of survey boundaries, inhomogeneous scans and atmospheric noise could degrade delensing efficiency if we simply adopt sub-optimal filtering to observed CMB data \cite{SPTpol:2017:delens}. However, efficient delensing would be realized by modifying the filtering appropriately. 

In this paper, we investigate impacts of observational systematic errors, which are unavoidable in practical situations, on the reconstruction of the lensing potential and evaluate the degradation of the delensing efficiency caused by propagation of the systematics through the mis-estimation of the lensing potential. 
Despite the multiple efforts on exploring the feasibility of delensing described above, impacts of observational systematics on delensing have not yet been thoroughly investigated for high-precision CMB polarization experiments. This paper, therefore, provides a first insight into the response of the delensed $B$-mode spectrum to several major observational systematics. 

This paper is organized as follows. The procedure of our simulations is described in Sec. \ref{method}. In Sec. \ref{reconstruction} and Sec. \ref{delensing}, we show results of the lensing reconstruction and subsequent delensing analysis, respectively. Finally, some discussion and our conclusion are in Sec. \ref{summary}.

\section{Simulation method} \label{method}
In early works of this research area, systematic errors are analytically modeled in terms of error fields which characterize how CMB signals deteriorate at each sky position and are classified according to their mathematical properties \cite{Hu2003,Shimon2008}. A couple of works followed in order to apply the models to the lensing reconstruction analysis \cite{Miller2009,Su2009}. Our study is in the line of such works. On the other hand, simulation-based assessments for evaluating the propagation of observational systematics in the lensing reconstruction analysis have been performed by several groups \cite{PB:phi:2019,Mirmelstein2020}. They adopted detailed setups of actual experiments (such as patch geometries, scan patterns, detector distributions, etc.) and carried out many sets of numerical simulations. Instead of performing simulations based on a specific experiment design, we take the former approach to obtain a correspondence between systematics and delensing efficiency in a robust and concrete form. 

We investigate three types of major systematic errors common in CMB polarization observations, i.e. gain error, angle error, and pointing error, the corrections of which still require conventional calibration procedures. Intensity (temperature) to polarization leakage, which was one of the most significant systematic errors, is avoided in principle by measurement with a polarization modulator \cite{Kusaka2014} though other types of systematic errors associated with the apparatus itself may arise instead \cite{DAlessandro2019,Duivenvoorden2020,Sekimoto2020}. Apart from them, beam-related mis-measurements such as elliptic deformation or sidelobe elongation are also major sources of systematic errors. Since the modelling of beam systematics in terms of error fields is still in development, we would like to leave this to future works.

We focus on the internal delensing analysis of the CMB polarization field in which we utilize statistics of the polarization field to reconstruct the lensing mass distribution. We adopt the well-established quadratic estimator method which is widely used in many works \cite{Okamoto2003}.

In the simulations discussed in this paper, we assume that polarization fields are noiseless and free from image blurring due to finite beam width for the purpose of clarifying how signal mis-measurements themselves affect the lensing analysis. In the subsequent sections \ref{reconstruction} and \ref{delensing}, we present results of our simulations in which a realization of the CMB polarization field is basically fixed and the systematic errors have various configurations in contrast to it. 

Throughout this paper, we adopt the fiducial $\Lambda$CDM parameter set based on the Planck 2018 baseline likelihood analysis \cite{Planck2018}. 
The values of the parameters are $h=0.6736$, $\Omega_b h^2=0.02237$, $\Omega_c h^2=0.12$, $\Omega_{\nu} h^2=0.000645$, $n_s=0.965$, $A_s(k_{\rm pivot}=0.05{\rm Mpc}^{-1})=2.099\times10^{-9}$, and $\tau=0.0544$. A single massive and two (effectively 2.046) massless neutrino species are assumed. 

\subsection{Systematic errors} \label{method_error}
A linear polarization state at a sky position $\hat{\mbox{\boldmath $n$}}$ is described in terms of Stokes parameters as $[Q \pm iU](\hat{\mbox{\boldmath $n$}})$. In this subsection, we describe how this expression is modified by imposing modulations due to the systematic errors on it.  

Uncertainty hosted in the gain, which is the correspondence of incident radiation to readout electrical signals, is a typical systematic error common among CMB polarization observations. Gain fluctuation caused by various instrumental and environmental factors during a scan observation results in residual calibration uncertainty which causes modulation of the polarization field. The polarization field modulated by the gain error is described as
\begin{equation}\label{syst-gain}
\begin{split}
(1+g(\hat{\mbox{\boldmath $n$}})) \, [Q \pm iU](\hat{\mbox{\boldmath $n$}}).
\end{split}
\end{equation}
The quantity $g(\hat{\mbox{\boldmath $n$}})$ is a gain bias at a sky position $\hat{\mbox{\boldmath $n$}}$. Note that a homogeneous bias gives just a scaling of the CMB field which is usually corrected by comparison with temperature or $E$-mode data provided by other observations.

Mis-estimation of polarization orientation angles causes false $B$-modes leaked from $E$-modes. The angle error is sourced from the thermal deformation and mechanical vibration of instruments, errors in telescope-attitude determination, and those in optical system calibration. Systematic effects of this kind are also caused by cosmic birefringence (e.g. \cite{Carroll:1989:rot,Minami2020}), which is considered to be induced by the interaction between CMB photons and some hypothetical field, or other mechanisms. The polarization field with the angle error is described as
\begin{equation}\label{syst-angle}
\begin{split}
e^{\pm 2i\alpha(\hat{\mbox{\boldmath $n$}})} \, [Q \pm iU](\hat{\mbox{\boldmath $n$}}).
\end{split}
\end{equation}
The quantity $\alpha(\hat{\mbox{\boldmath $n$}})$ is an angle bias at a sky position $\hat{\mbox{\boldmath $n$}}$. Here, we don't restrict the angle error to have spatial homogeneity which is often assumed in the literature. 

A mismatch between an actual pointing direction and its corresponding estimated direction degrades the spatial position accuracy in a polarization map. Pattern distortion due to the pointing error modulates the polarization field and converts the associated parity modes from $E$ to $B$ (or vice versa) in a similar manner to the gravitational lensing effects. The pointing error shares much of its physical origin with the angle error. The modulated polarization field is described as
\begin{equation}\label{syst-pointing}
\begin{split}
[Q \pm iU](\hat{\mbox{\boldmath $n$}}+d(\hat{\mbox{\boldmath $n$}})).
\end{split}
\end{equation}
The vector $d(\hat{\mbox{\boldmath $n$}})$ stands for the discrepancy between an actually observed sky position and a corresponding mapped position $\hat{\mbox{\boldmath $n$}}$ due to the pointing error. It is generally decomposed into two components \cite{Hirata2003,Cooray2005,Namikawa2012} as
\begin{equation}\label{pointing-type}
\begin{split}
d = \nabla \psi  + (\star \nabla) \varpi . 
\end{split}
\end{equation}
$\psi$ and $\varpi$ are potential functions of gradient (even parity) and curl (odd parity) modes, respectively. The $\star$ operator denotes counterclockwise rotation by 90 degrees. The curl component in this case is often non-negligible unlike in the case of the  gravitational lensing effects. In fact, a scan orbit of a telescope boresight is often something like a periodic curve and a static pointing bias tends to make the error vectors swirly distributed on the sky. Also, random errors don't prefer one of the two components. It has been reported that fluctuation of sampling positions due to pointing errors causes smearing of the small-scale inhomogeneity of the polarization fields as an additional effect \cite{Mirmelstein2020,QUIET2012}. That effect is not described by this error model. In those works, it is modeled as an additional suppression factor applied to multipole moments of the CMB polarization. The suppression factor is parameterized by an effective beam width which stands for a length scale of the smearing. We just comment that the overall amplitude of a lensing potential reconstructed in the presence of the suppression factor tends to be reduced, and consequently the delensing efficiency is degraded, though the response is not very sensitive to the amplitude reduction. We would like to report on this issue in a future further study.

$g(\hat{\mbox{\boldmath $n$}})$, $\alpha(\hat{\mbox{\boldmath $n$}})$, and $d(\hat{\mbox{\boldmath $n$}})$ stand for net errors of gain, angle, and pointing at a sky position $\hat{\mbox{\boldmath $n$}}$, respectively. A net error at a sky position is an average of the instantaneous errors, each of which is associated with a single piece of data sampled at the position during a scan observation. Since each sky position is visited many times during an entire observation period, the net error is averaged out in a statistical manner. For example, if the fluctuation time scale of an error is shorter than the revisiting intervals, the net amplitude of the error is suppressed by a factor of 1/$\sqrt{N}$, where $N$ is the number of visiting times. In most contemporary experiments, the net amplitude of an error is smaller by one or more orders than the instantaneous amplitude of the error. 

In a single simulation case, the field of the systematic error ($g(\hat{\mbox{\boldmath $n$}})$, $\alpha(\hat{\mbox{\boldmath $n$}})$, or $d(\hat{\mbox{\boldmath $n$}})$) is set to have a non-vanishing value of the multipole moment only at a single multipole ($\ell=\ell_{\rm in}$) so that we can clearly see the scale dependence of the analysis results. (See also Appendix \ref{eqs-EBlm}.) Then, given the statistical isotropy of the CMB polarization field, we set its directional eigenvalue fixed to zero ($m=0$). We normalize the overall amplitude of the systematic error by quantifying a spatial root-mean-square (RMS) value of the error field. 

We utilize routines of the {\tt Healpix}\footnote{https://healpix.jpl.nasa.gov/} library to manipulate the numerical data of the polarization and error fields and {\tt Lenspix}\footnote{http://cosmologist.info/lenspix/} to make lensed CMB polarization maps on which to impose the systematic errors. 

\subsection{Lensing analysis} \label{method_analysis}
To reconstruct the lensing potential, we adopt the $EB$ quadratic estimator \cite{Okamoto2003}, which has decisive reconstruction performance in the situation concerned here, given as
\begin{equation}\label{EB-est}
\begin{split}
 [\widehat{\phi}^{EB}_{LM}]^* = A^{EB}_L\sum_{\ell\ell'}\sum_{mm'}
  \left(
   \begin{array}{ccc}
    \ell & \ell' & L \\
    m & m' & M \\
   \end{array}
  \right) 
 g^{EB}_{\ell\ell'L}\widehat{E}_{\ell m}\widehat{B}_{\ell'm'} \, .
\end{split}
\end{equation}
$\widehat{E}_{\ell m}$ and $\widehat{B}_{\ell'm'}$  are observed harmonic coefficients of $E$- and $B$-modes, respectively. Hereafter in this subsection , we denote the quantities to be affected by the systematic errors as characters with  \, $\widehat{}$ \,  while quantities theoretically predetermined are without such symbols. We define the quantities $g^{EB}_{\ell\ell'L}$ and $A^{EB}_L$ as
\begin{equation}\label{EB-g-A}
\begin{split}
 g^{EB}_{\ell\ell'L} &= \frac{[f^{EB}_{\ell\ell'L}]^*}{C_{\ell}^{EE}C_{\ell'}^{BB}} , \\ 
 A^{EB}_L &= \left\{ \frac{1}{2L+1}\sum_{\ell\ell'}f^{EB}_{\ell\ell'L}g^{EB}_{\ell\ell'L} \right\}^{-1} .
\end{split}
\end{equation}
$C_{\ell}^{EE}$ and $C_{\ell}^{BB}$ are the angular power spectra of the lensed $E$- and $B$-modes, respectively. 
The weight function $f^{EB}_{\ell\ell'L}$, which is defined as
\begin{equation}\label{EB-f}
\begin{split}
f^{EB}_{\ell\ell'L} = i C_{\ell}^{EE} S_{\ell' \ell L}^{(-)}, 
\end{split}
\end{equation}
is the combination of the lensed $E$-mode spectrum \cite{Lewis2011,Anderes2013} and the mode coupling function $(S_{\ell' \ell L}^{(-)})$, which is  given by
\begin{equation}\label{EB-S}
\begin{split}
 S^{(-)}_{\ell\ell'L} &= \frac{1- (-1)^{\ell+\ell'+L}}{2} \sqrt{\frac{(2\ell+1)(2\ell'+1)(2L+1)}{16\pi}} \\ 
 &\qquad \times [-\ell(\ell+1)+\ell'(\ell'+1)+L(L+1)]
  \left(
   \begin{array}{ccc}
    \ell & \ell' & L \\
    2 & -2 & 0 \\
   \end{array}
  \right) .
\end{split}
\end{equation}

By use of the reconstructed lensing potential, we make a template of the lensing $B$-modes for delensing by convolving the $E$-modes with the reconstructed lensing potential as 
\begin{equation}\label{B-temp}
\begin{split}
 \widehat{B}^{\rm lens}_{\ell m} &= -i \sum_{\ell'm'}\sum_{LM}
  \left(
   \begin{array}{ccc}
    \ell & \ell' & L \\
    m & m' & M \\
   \end{array}
  \right)
 S^{(-)}_{\ell\ell'L} W^{E}_{\ell'} W^{\phi}_L \widehat{E}_{\ell'm'}^{*}\widehat{\phi}_{LM}^{*} \, ,
\end{split}
\end{equation}
where the Wiener filters are defined as
\begin{equation}\label{Wiener}
\begin{split}
W^{\phi}_L &= \frac{ C^{\phi\phi}_L }{ C^{\phi\phi}_L+A_{L}^{EB} } , \\
W^{E}_{\ell} &= 1.
\end{split}
\end{equation}
We don't apply filtering to the $E$-modes because noise is absent in our case.
Residual $B$-modes after delensing are evaluated as 
\begin{equation}\label{B-residual}
\begin{split}
 \widehat{B}^{\rm res}_{\ell m} = B_{\ell m} - \widehat{B}^{\rm lens}_{\ell m} .
\end{split}
\end{equation}
To estimate the delensing efficiency itself, we set the first term of the right-hand side of the equation above to be a multipole moment of the lensing $B$-modes which is initially prepared and does not include biases from the systematic errors, except for the case of the gradient-type pointing error. (See Sec. \ref{rec_pointing} and \ref{del_pointing} for details.). 
The lensing template described above does not include higher-order terms of the lensing potential, but the power spectrum of the delensed $B$-modes does not have a significant contribution from $\mC{O}(\phi^2)$ terms because of the partial cancellation as shown in Ref. \cite{BaleatoLizancos:2020b}. 

It is reported that the overlap of multipole ranges used for reconstruction and delensing causes an undesirable bias in the residual lensing $B$-modes defined above \cite{Namikawa2014a,Lizancos:2020:delens-bias}. We incorporate only multipoles between $\ell=300$ and $2048$ into the reconstruction analysis to avoid such delensing bias and evaluate the residual lensing B-modes only between $\ell=2$ and $299$. 

We compute the theoretical angular power spectra of the CMB polarization and lensing potential with {\tt CAMB}\footnote{https://camb.info/}. The library used for the lensing analysis described above is made public at {\tt GitHub}\footnote{https://github.com/toshiyan/}.

\section{Lensing reconstruction} \label{reconstruction}
Contributions from the systematic errors to polarization fields are not in a state of pure even parity which is kept by the primary CMB polarization of the density fluctuation origin. In fact, $E$- and $B$-modes induced by the systematic errors have similar amplitude. 
Considering the fact that the lensing $B$-modes are smaller than the CMB $E$-modes by a few orders, we contrast the systematics induced $B$-modes to the lensing $B$-modes to illustrate impacts on the $EB$ quadratic estimator. In this section, we compare angular power spectra of the induced $B$-modes with reconstructed lensing potentials and discuss the response of the reconstructed potentials to the systematic errors imposed to the CMB polarization field. 

We evaluate the angular power spectra of the reconstructed lensing potentials as \cite{Namikawa2014b}
\begin{equation}\label{Clpp}
\begin{split}
 \widehat{C}^{\phi\phi}_L = \frac{1}{2L+1}\sum_{M=-L}^{L}|\widehat{\phi}_{LM}|^2 - N^{(0)}_L ,
\end{split}
\end{equation}
where the Gaussian bias term is given by $N^{(0)}_L = A_L$ in the full-sky idealistic case \cite{Okamoto2003}. We simply use the normalization factor to evaluate the Gaussian bias though the realization-dependent method \cite{Namikawa:2012:bhe} is more effective to mitigate systematics in $N^{(0)}_L$ which are often present in non-idealistic cases \cite{Mirmelstein2020}. Although we can evaluate the $N^{(1)}$ bias contribution to the spectrum by means of an analytic calculation or a Monte Carlo simulation \cite{P13:phi}, we just omit the correction. These treatments do not affect our discussion in this paper. 

\subsection{Gain error} \label{rec_gain}

\begin{figure*}
\hspace{-13pt}
\subfloat[]{
\includegraphics[width=3.05in]{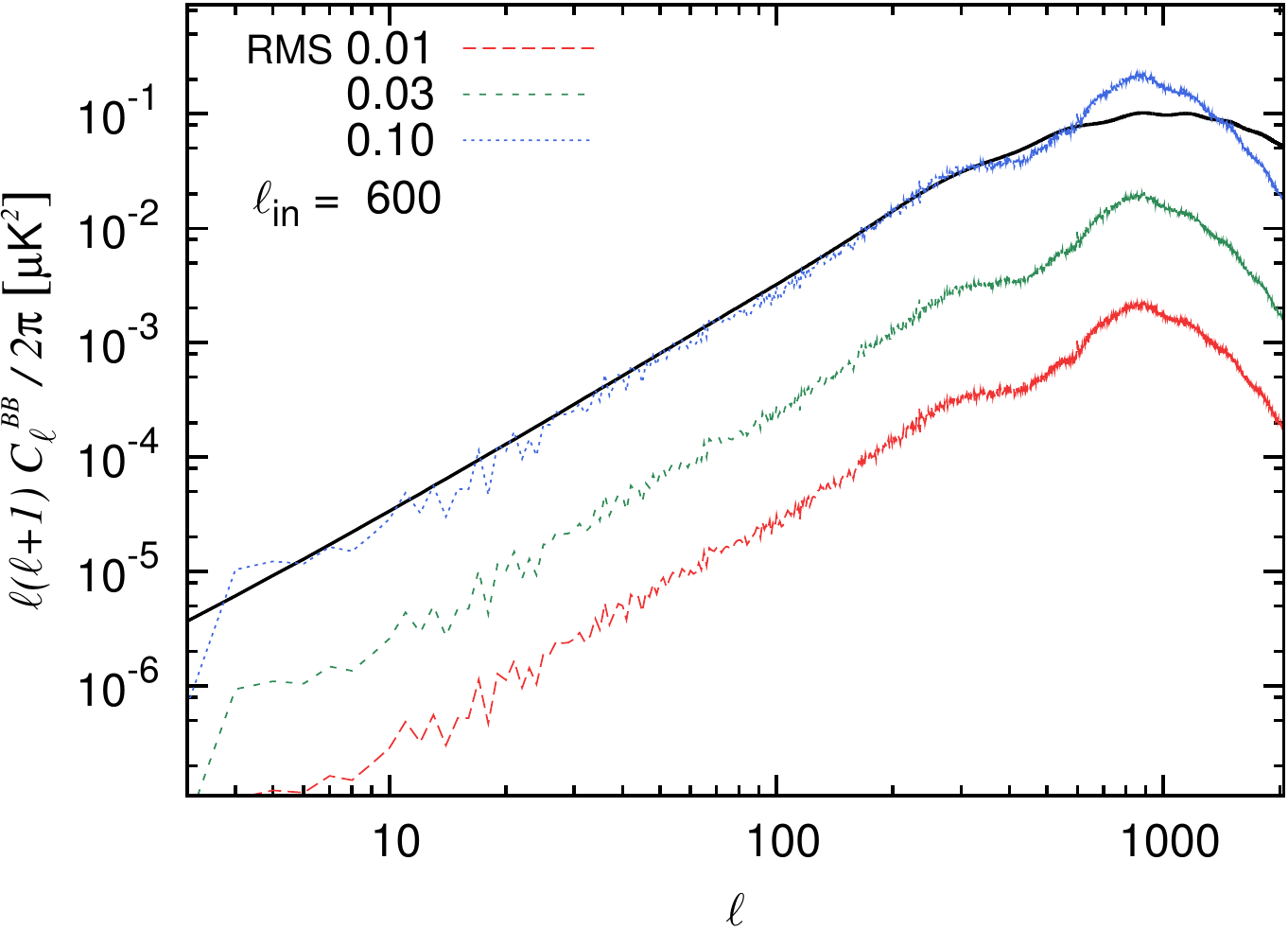}\label{clbb_sys_gain_a} }
\hspace{-4pt}
\subfloat[]{ 
\includegraphics[width=3.05in]{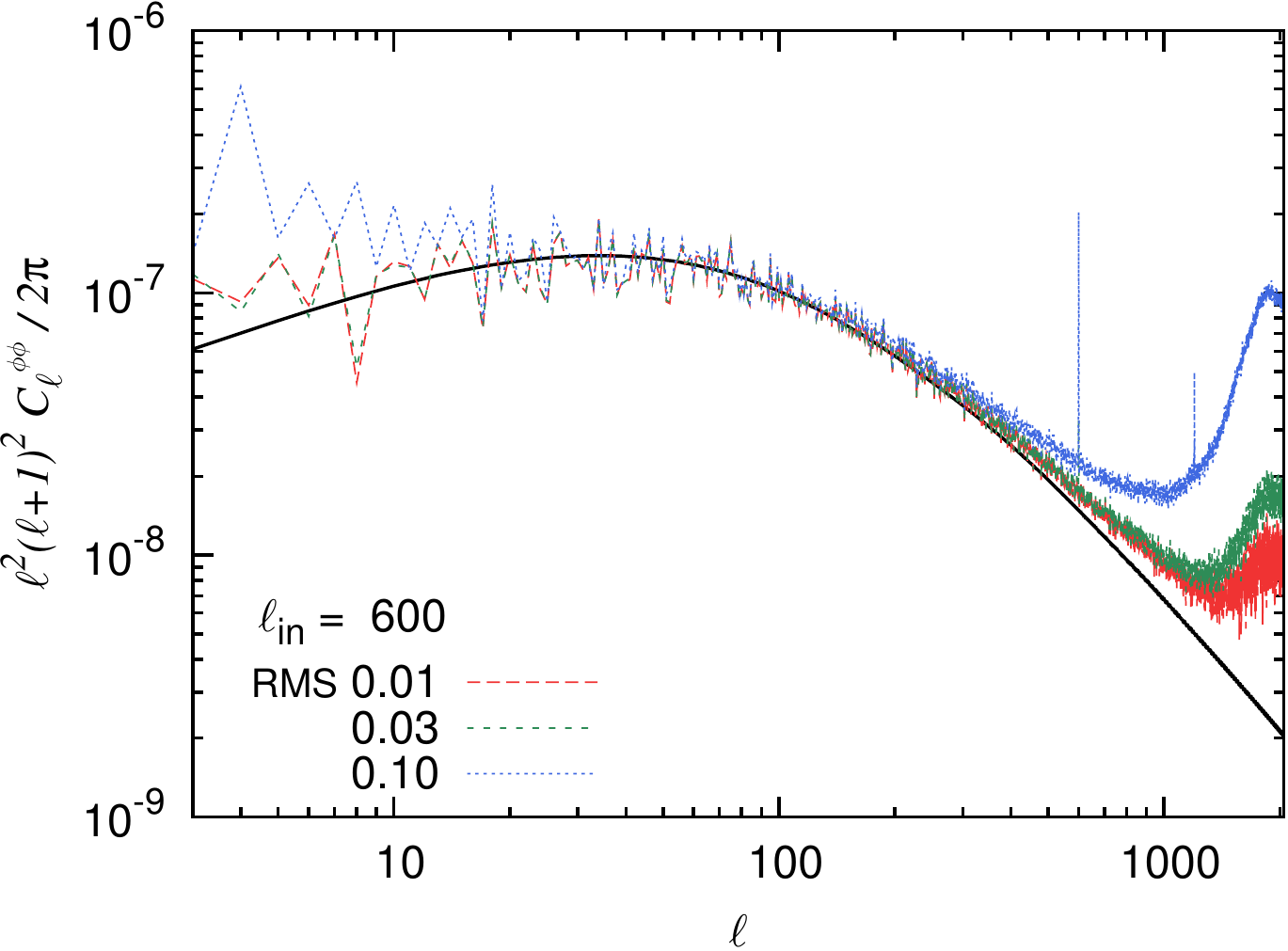}\label{clpp_rec_gain_a} }
\caption{Dependence on RMS amplitude of the gain error. {\it Left}: angular power spectra of $B$-modes induced by the gain error. The thick black curve is the theoretical spectrum of the lensing $B$-modes. {\it Right}: angular power spectra of lensing potentials reconstructed in presence of the gain error. The thick black curve is the theoretical spectrum of the lensing potential.}
\label{gain_a}
\end{figure*}

\begin{figure*}
\subfloat[]{
\hspace{-13pt}
\includegraphics[width=3.05in]{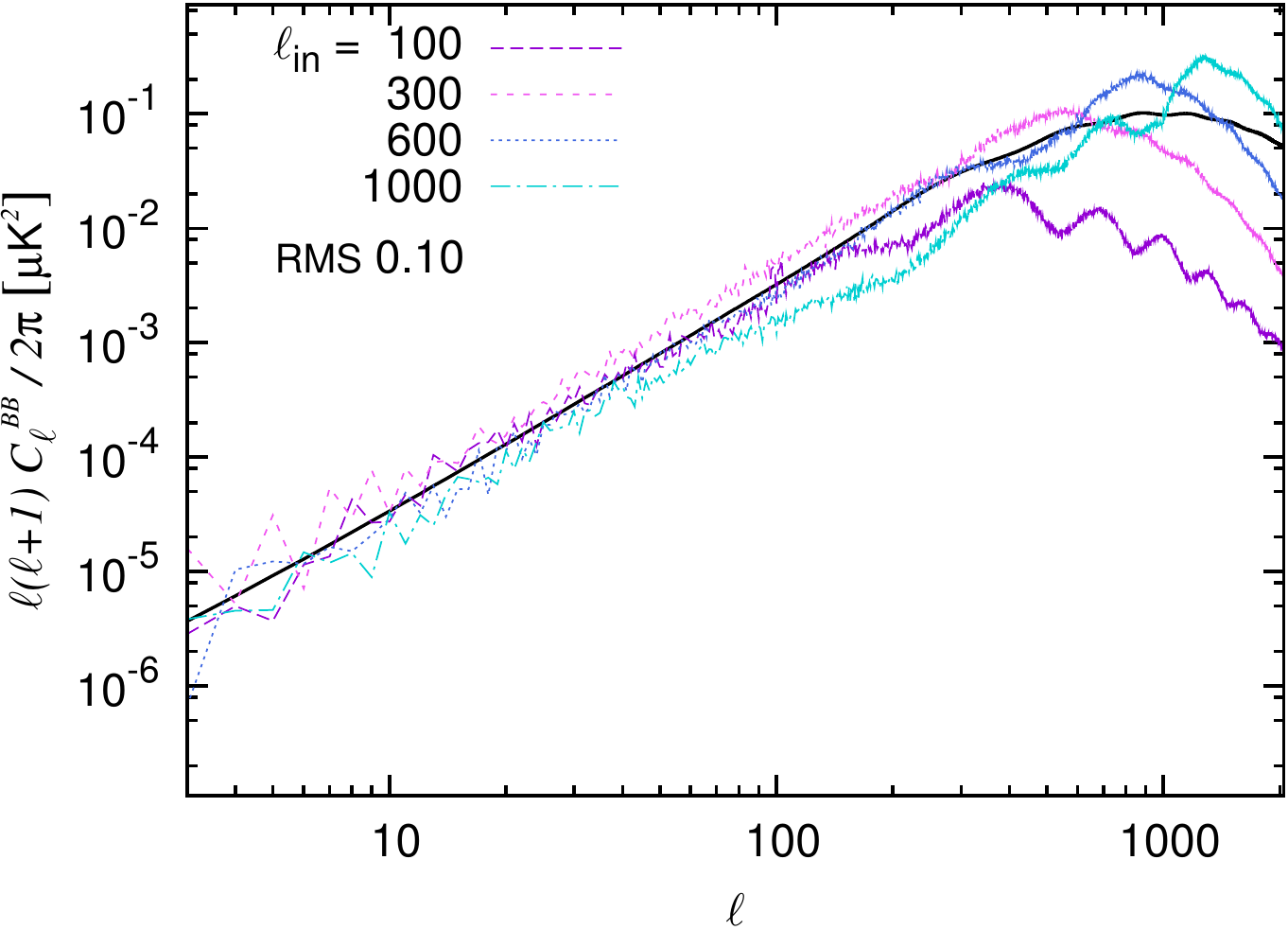}\label{clbb_sys_gain_s} }
\hspace{-4pt}
\subfloat[]{
\includegraphics[width=3.05in]{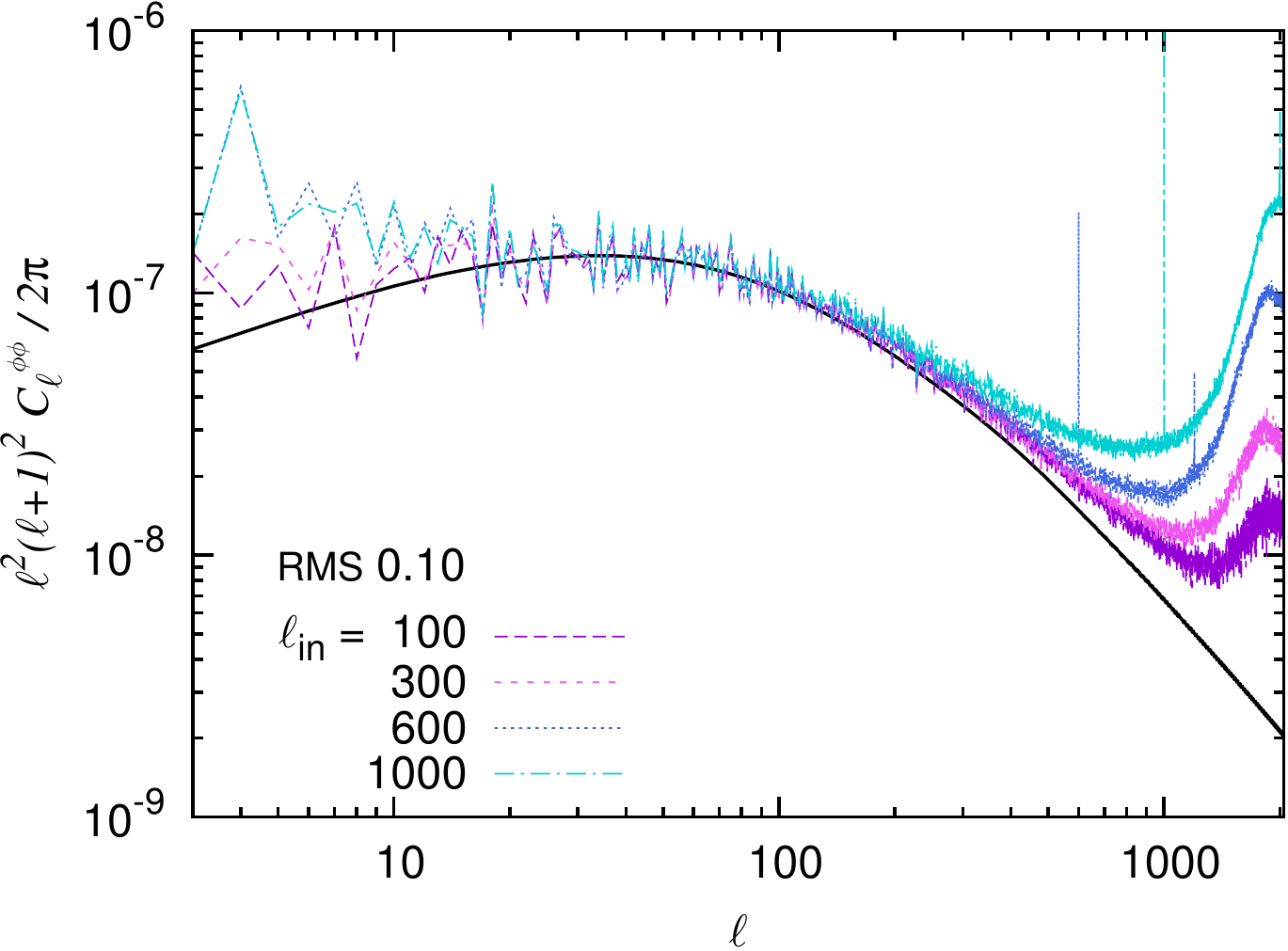}\label{clpp_rec_gain_s} }
\caption{Dependence on modulation scale ($\ell_{\rm in}$) of the gain error. {\it Left}: angular power spectra of $B$-modes induced by the gain error. The thick black curve is the theoretical spectrum of the lensing $B$-modes. {\it Right}: angular power spectra of lensing potentials reconstructed in presence of the gain error. The thick black curve is the theoretical spectrum of the lensing potential.}
\label{gain_s}
\end{figure*}

The figure \ref{clbb_sys_gain_a} shows the angular power spectra of $B$-modes induced by the gain error with $\ell_{\rm in} = 600$. The RMS values of the error fields are $0.01$, $0.03$, and $0.10$. The figure \ref{clpp_rec_gain_a} shows the angular power spectra of the reconstructed lensing potentials in the respective cases. First, let us mention the overall appearance of the figures. While the shapes of the $B$-mode spectra are apparently similar to that of the lensing $B$-modes, the spectra of the reconstructed lensing potentials have spiky features at the multipoles of the imposed error field and its overtones, which is clearly seen in the case of RMS $0.10$. This gives us a simple intuition that multipole moments of reconstructed lensing potentials exhibit such a sharp response to the power of error fields at corresponding multipoles. Also, it is found that a trend of deviation from the theoretical spectrum of the lensing potential is observed both in the scale ranges of several dozen degrees and sub-degree. The bias in the latter range possibly contaminates the delensing analysis, which is discussed in detail in the sections \ref{rec_mf} and \ref{delensing}. Next, we discuss the dependence of reconstructed lensing potentials on the RMS values of fields of the gain error. In the case of RMS $0.01$, the amplitude of the induced $B$-mode spectrum is about $1\%$ of that of the lensing $B$-mode spectrum, and the reconstructed lensing potential is almost identical to the one ideally reconstructed in the absence of any systematic errors. When we increase the RMS value from $0.01$, the spectrum of the reconstructed lensing potential begins to deviate perceptibly from the ideally reconstructed spectrum around RMS $0.03$ in the case of which the amplitude of the induced $B$-mode spectrum is about $10\%$ of that of the lensing $B$-mode spectrum. 

The figure \ref{clbb_sys_gain_s} shows the angular power spectra of $B$-modes induced by the gain error the fields of which have their RMS values fixed at $0.10$. The adopted multipoles of the error fields are $100$, $300$, $600$, and $1000$. The figure \ref{clpp_rec_gain_s} shows the angular power spectra of the reconstructed lensing potentials in the respective cases. When we increase $\ell_{\rm in}$, the induced $B$-modes grow substantially, especially in small scales. The multipoles of the spikes in each of the reconstructed lensing potentials are consistent with the adopted multipole of the corresponding case. The coherent deviation from the theoretical spectrum of the lensing potential in the scale ranges of several dozen degrees and sub-degree shows response to the amplitude of the induced $B$-modes in small scales. 

\subsection{Angle error} \label{rec_angle}

\begin{figure*}
\hspace{-13pt}
\subfloat[]{
\includegraphics[width=3.05in]{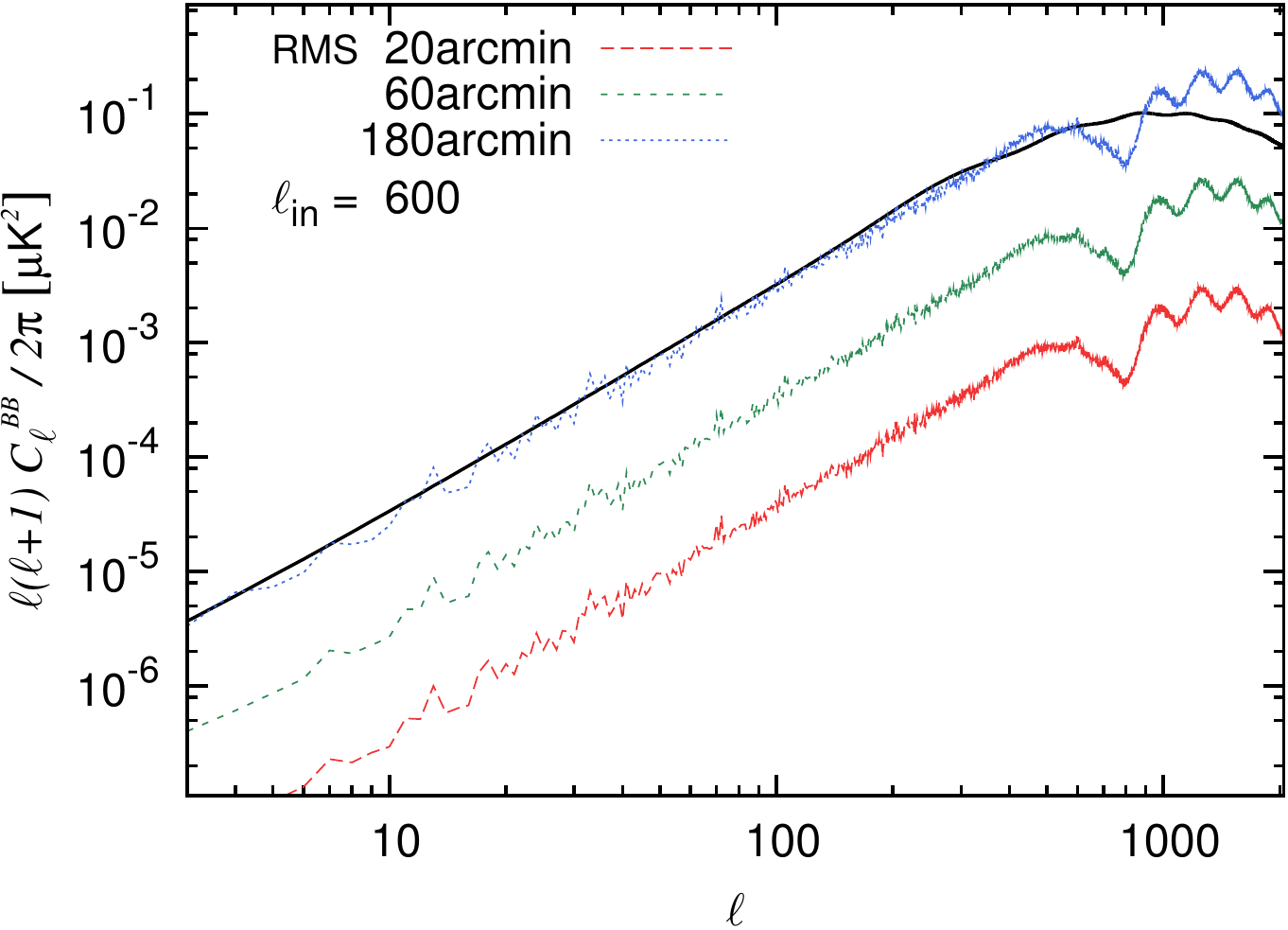}\label{clbb_sys_angl_a} }
\hspace{-4pt}
\subfloat[]{
\includegraphics[width=3.05in]{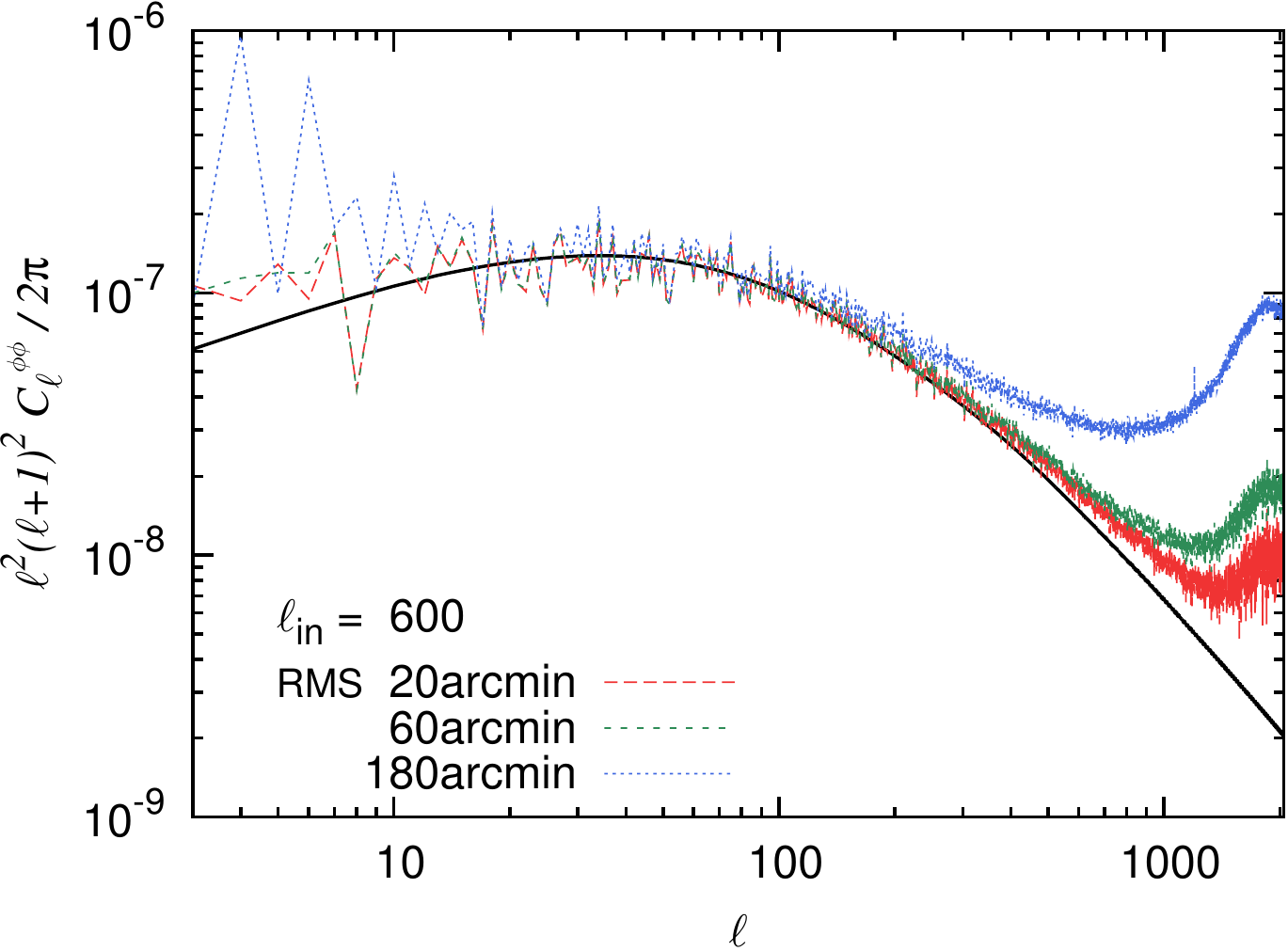}\label{clpp_rec_angl_a} }
\caption{Dependence on RMS amplitude of the angle error. {\it Left}: angular power spectra of $B$-modes induced by the angle error. The thick black curve is the theoretical spectrum of the lensing $B$-modes. {\it Right}: angular power spectra of lensing potentials reconstructed in presence of the angle error. The thick black curve is the theoretical spectrum of the lensing potential.}
\label{angl_a}
\end{figure*}

\begin{figure*}
\hspace{-13pt}
\subfloat[]{
\includegraphics[width=3.05in]{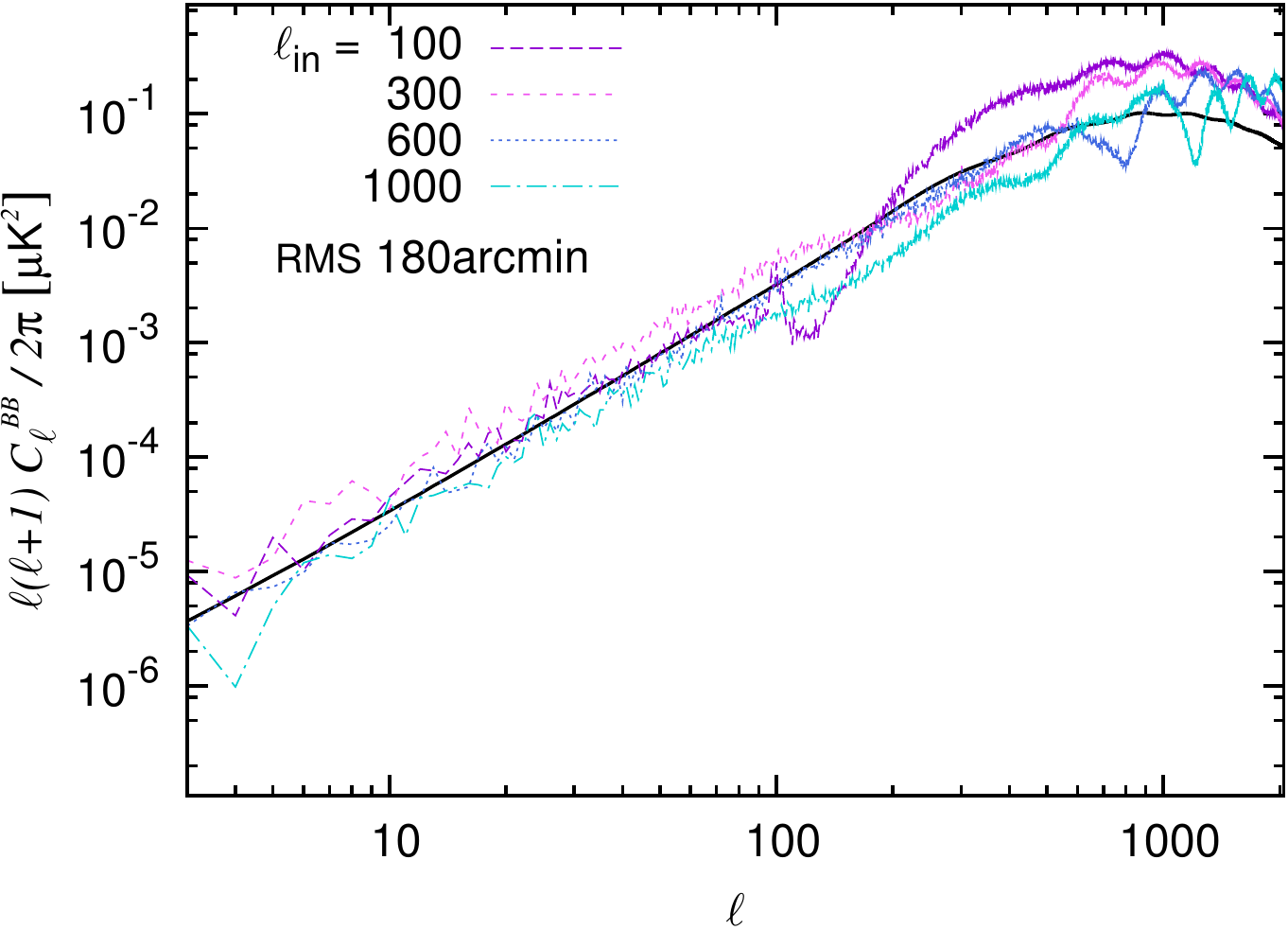}\label{clbb_sys_angl_sh} }
\hspace{-4pt}
\subfloat[]{
\includegraphics[width=3.05in]{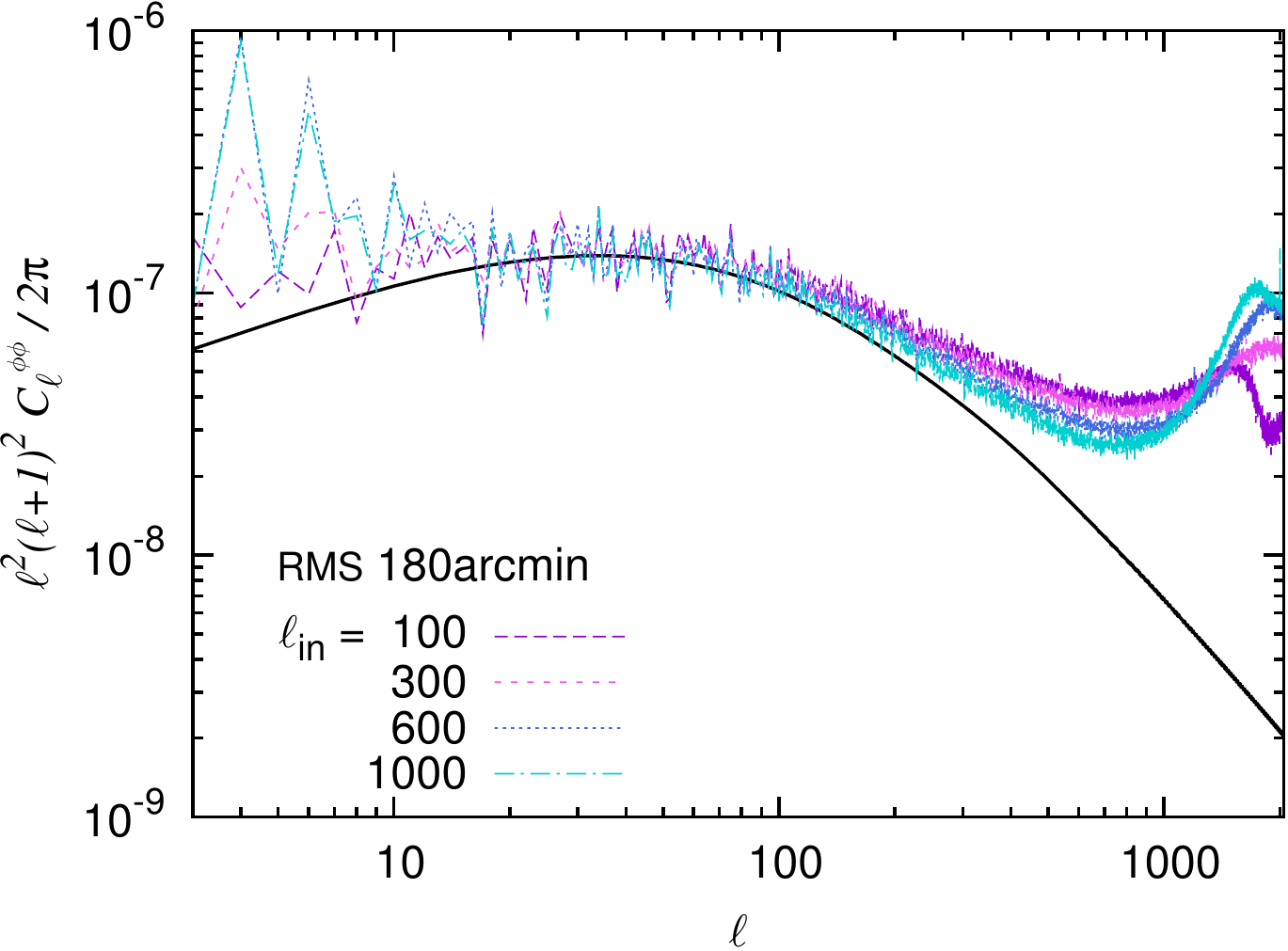}\label{clpp_rec_angl_sh} }
\caption{Dependence on modulation scale ($\ell_{\rm in}$) of the angle error. {\it Left}: angular power spectra of $B$-modes induced by the angle error. The thick black curve is the theoretical spectrum of the lensing $B$-modes. {\it Right}: angular power spectra of lensing potentials reconstructed in presence of the angle error. The thick black curve is the theoretical spectrum of the lensing potential.}
\label{angl_sh}
\end{figure*}

The figure \ref{clbb_sys_angl_a} shows the angular power spectra of $B$-modes induced by the angle error with $\ell_{\rm in} = 600$. The RMS values of the error fields are $20$\,arcmin, $60$\,arcmin, and $180$\,arcmin. The figure \ref{clpp_rec_angl_a} shows the angular power spectra of the reconstructed lensing potentials in the respective cases. The induced $B$-modes and the reconstructed lensing potentials exhibit similar behavior to those shown in the figure \ref{gain_a} for the gain error, except for the spike which is observed only at the overtone scale of the adopted multipole (see also Fig. \ref{clpp_rec_angl_mf}). 

The figure \ref{clbb_sys_angl_sh} shows the angular power spectra of $B$-modes induced by the angle error the fields of which have their RMS values fixed at $180$\,arcmin. The adopted multipoles of the error fields are $100$, $300$, $600$, and $1000$. The figure \ref{clpp_rec_angl_sh} shows the angular power spectra of the reconstructed lensing potentials in the respective cases. The induced $B$-mode spectra appear to be overlapping each other. The spectra of the reconstructed lensing potentials are also similar to each other. It is well known that a spectral shape of false $B$-modes leaked from $E$-modes due to a spatially homogeneous angular bias is identical to that of the $E$-modes. In fact, when we further decrease $\ell_{\rm in}$ from $100$, it is found that the spectral shape of the induced $B$-modes approaches that of the CMB $E$-modes and the induced $B$-mode spectrum exhibits prominent oscillatory structure which comes from the acoustic peaks seen in the CMB $E$-mode spectrum. In that case, the overall amplitude of the induced $B$-mode spectrum does not significantly change and the reconstructed lensing potential keeps its spectrum almost identical to that in the case of $\ell_{\rm in}=100$. 

\subsection{Pointing error} \label{rec_pointing}

\begin{figure*}
\hspace{-13pt}
\subfloat[]{
\includegraphics[width=3.05in]{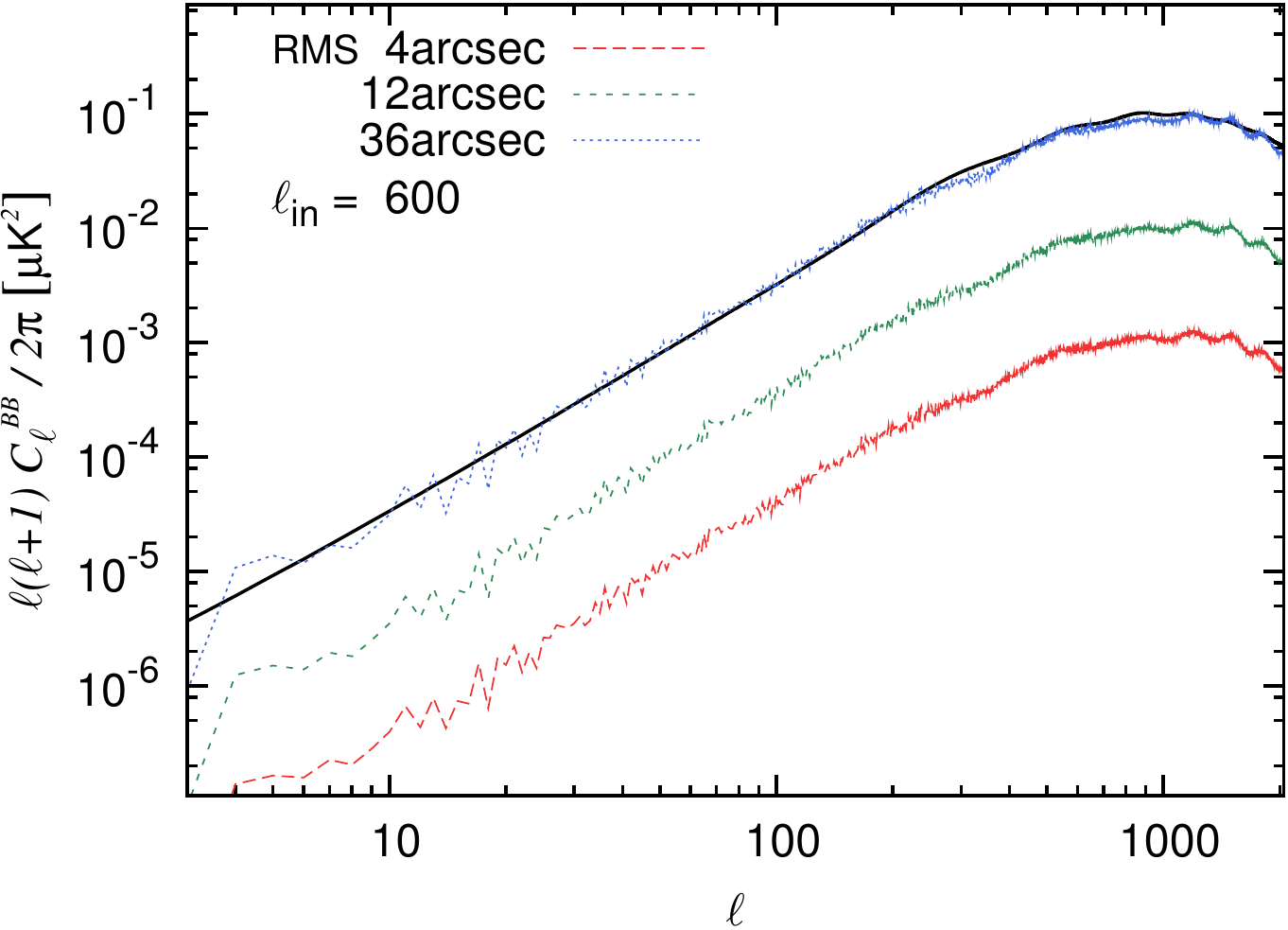}\label{clbb_sys_grad_a} }
\hspace{-4pt}
\subfloat[]{
\includegraphics[width=3.05in]{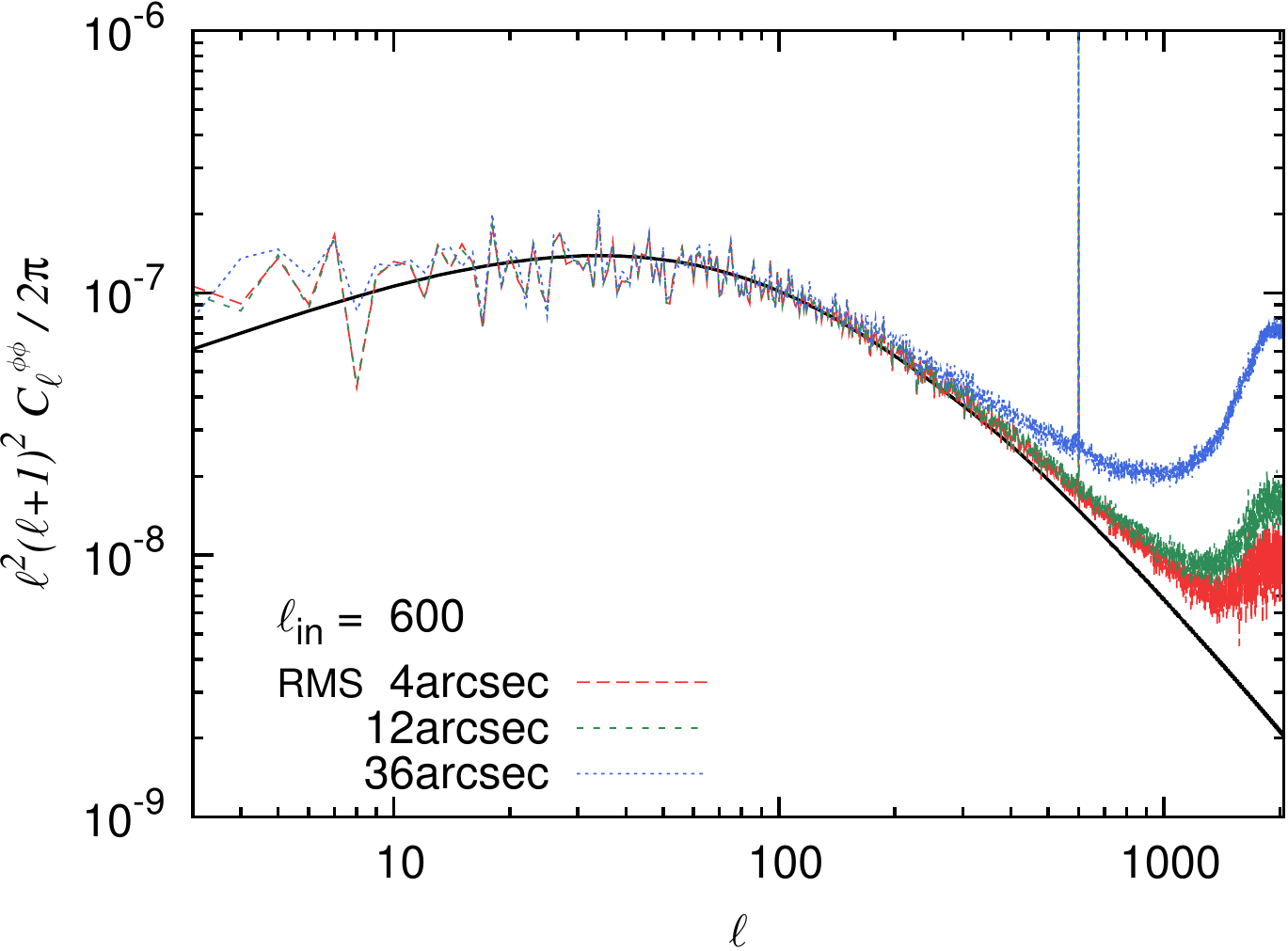}\label{clpp_rec_grad_a} }
\caption{Dependence on RMS amplitude of the gradient-type pointing error. {\it Left}: angular power spectra of $B$-modes induced by the gradient-type pointing error. The thick black curve is the theoretical spectrum of the lensing $B$-modes. {\it Right}: angular power spectra of lensing potentials reconstructed in presence of the gradient-type pointing error. The thick black curve is the theoretical spectrum of the lensing potential.}
\label{grad_a}
\end{figure*}

\begin{figure*}
\hspace{-13pt}
\subfloat[]{
\includegraphics[width=3.05in]{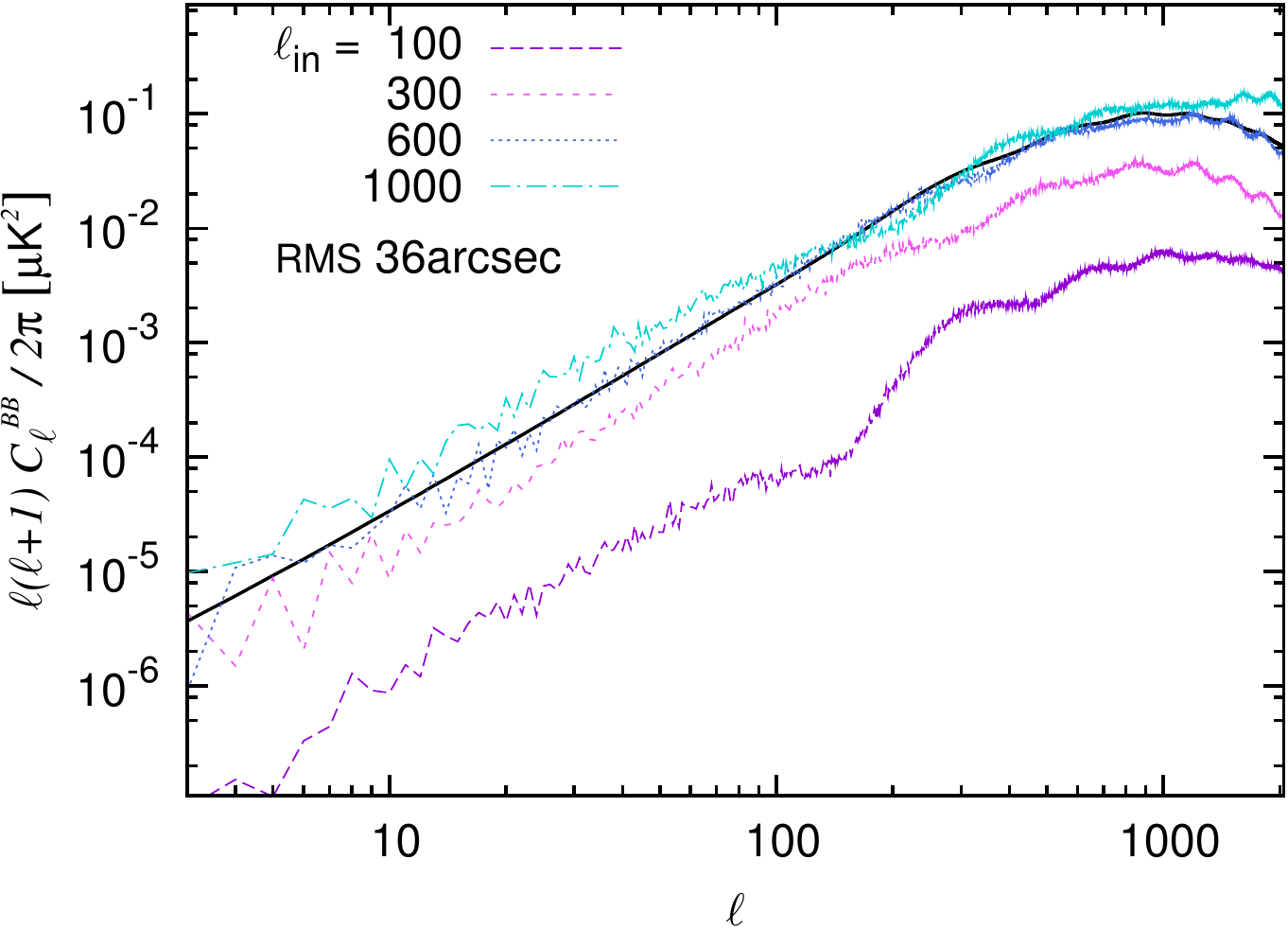}\label{clbb_sys_grad_s} }
\hspace{-4pt}
\subfloat[]{
\includegraphics[width=3.05in]{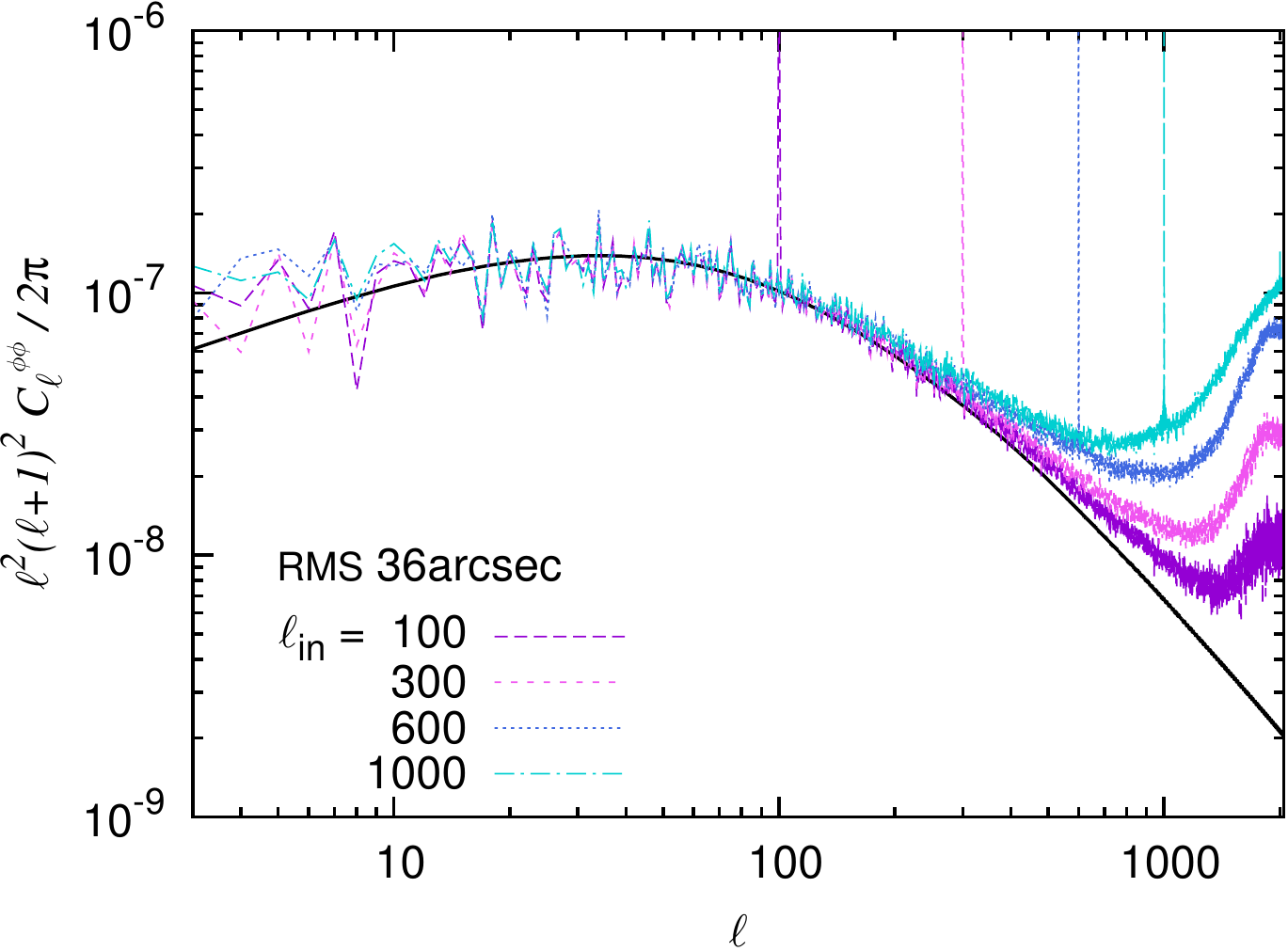}\label{clpp_rec_grad_s} }
\caption{Dependence on modulation scale ($\ell_{\rm in}$) of the gradient-type pointing error. {\it Left}: angular power spectra of $B$-modes induced by the gradient-type pointing error. The thick black curve is the theoretical spectrum of the lensing $B$-modes. {\it Right}: angular power spectra of lensing potentials reconstructed in presence of the gradient-type pointing error. The thick black curve is the theoretical spectrum of the lensing potential.}
\label{grad_s}
\end{figure*}

\begin{figure*}
\hspace{-13pt}
\subfloat[]{
\includegraphics[width=3.05in]{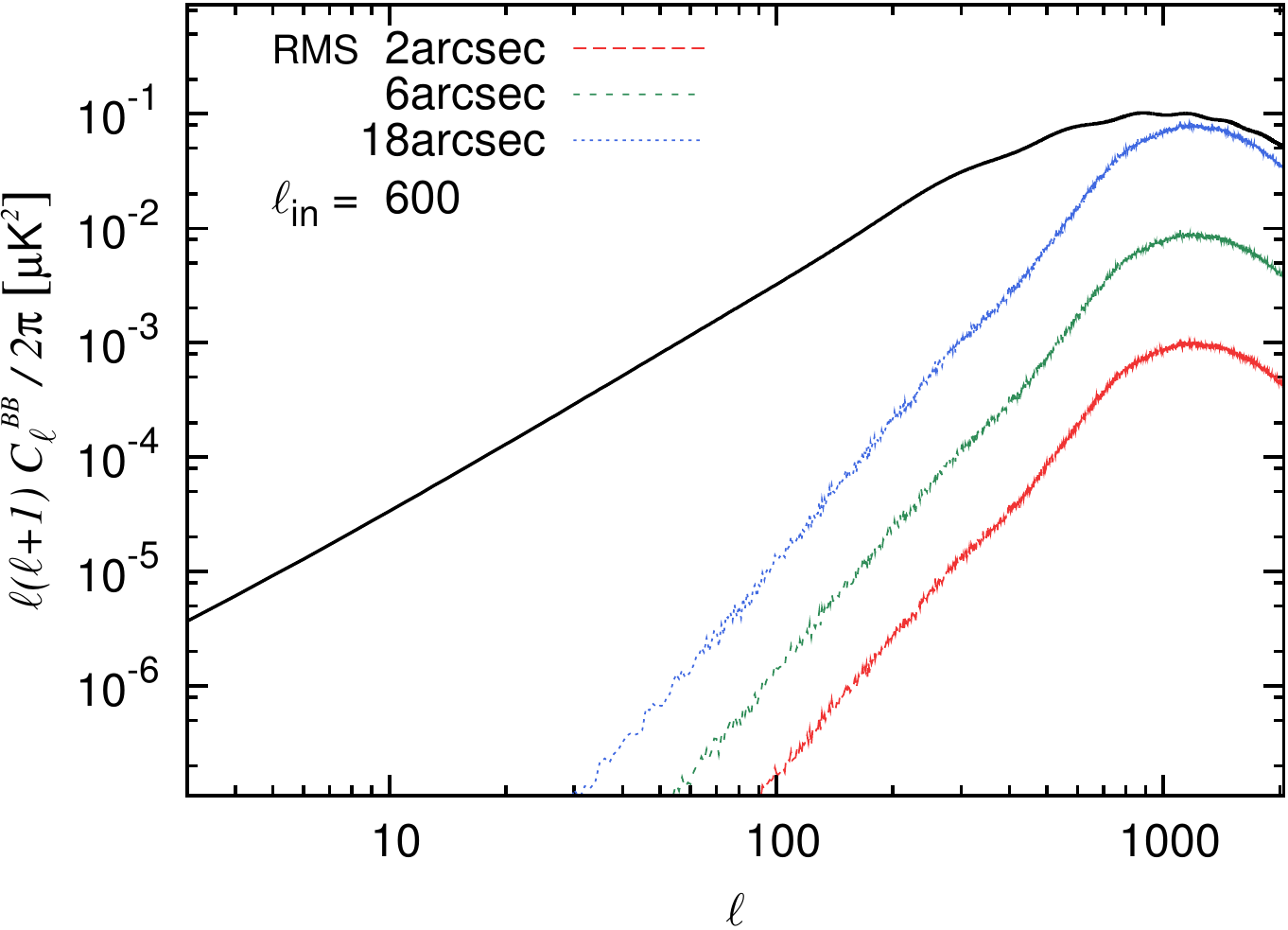}\label{clbb_sys_curl_a} }
\hspace{-4pt}
\subfloat[]{
\includegraphics[width=3.05in]{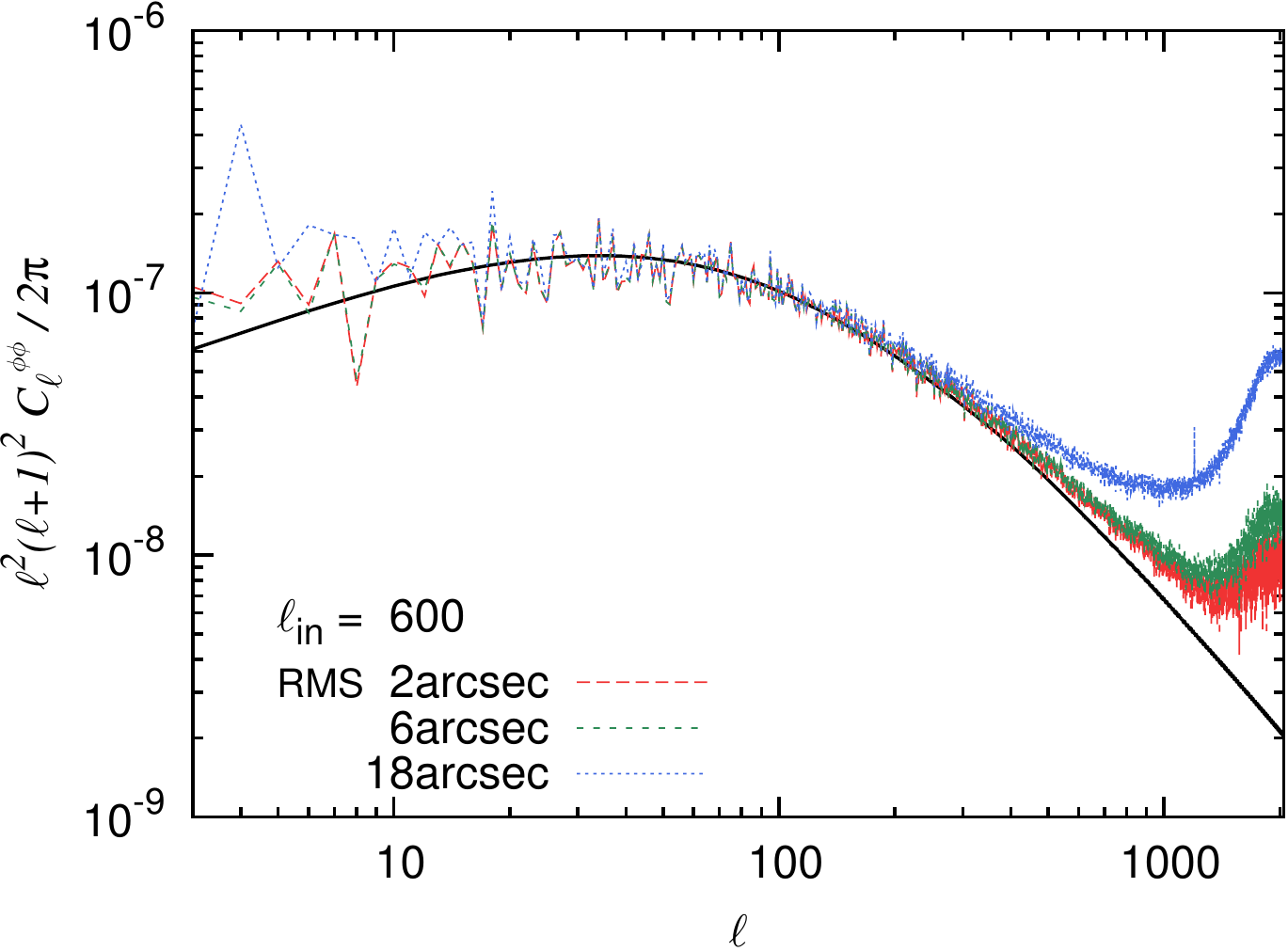}\label{clpp_rec_curl_a} }
\caption{Dependence on RMS amplitude of the curl-type pointing error. {\it Left}: angular power spectra of $B$-modes induced by the curl-type pointing error. The thick black curve is the theoretical spectrum of the lensing $B$-modes. {\it Right}: angular power spectra of lensing potentials reconstructed in presence of the curl-type pointing error. The thick black curve is the theoretical spectrum of the lensing potential.}
\label{curl_a}
\end{figure*}

\begin{figure*}
\hspace{-13pt}
\subfloat[]{
\includegraphics[width=3.05in]{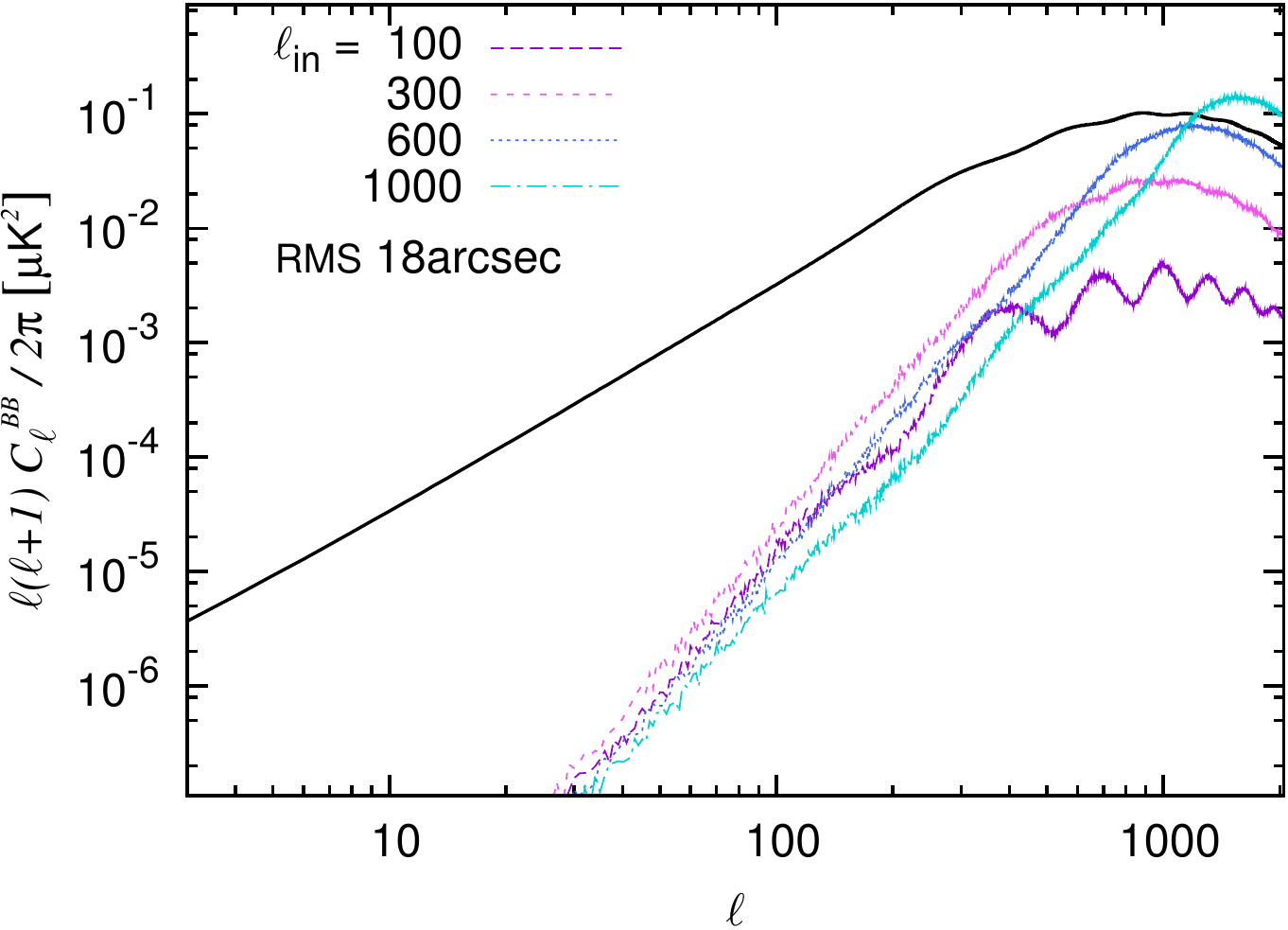}\label{clbb_sys_curl_s} }
\hspace{-4pt}
\subfloat[]{
\includegraphics[width=3.05in]{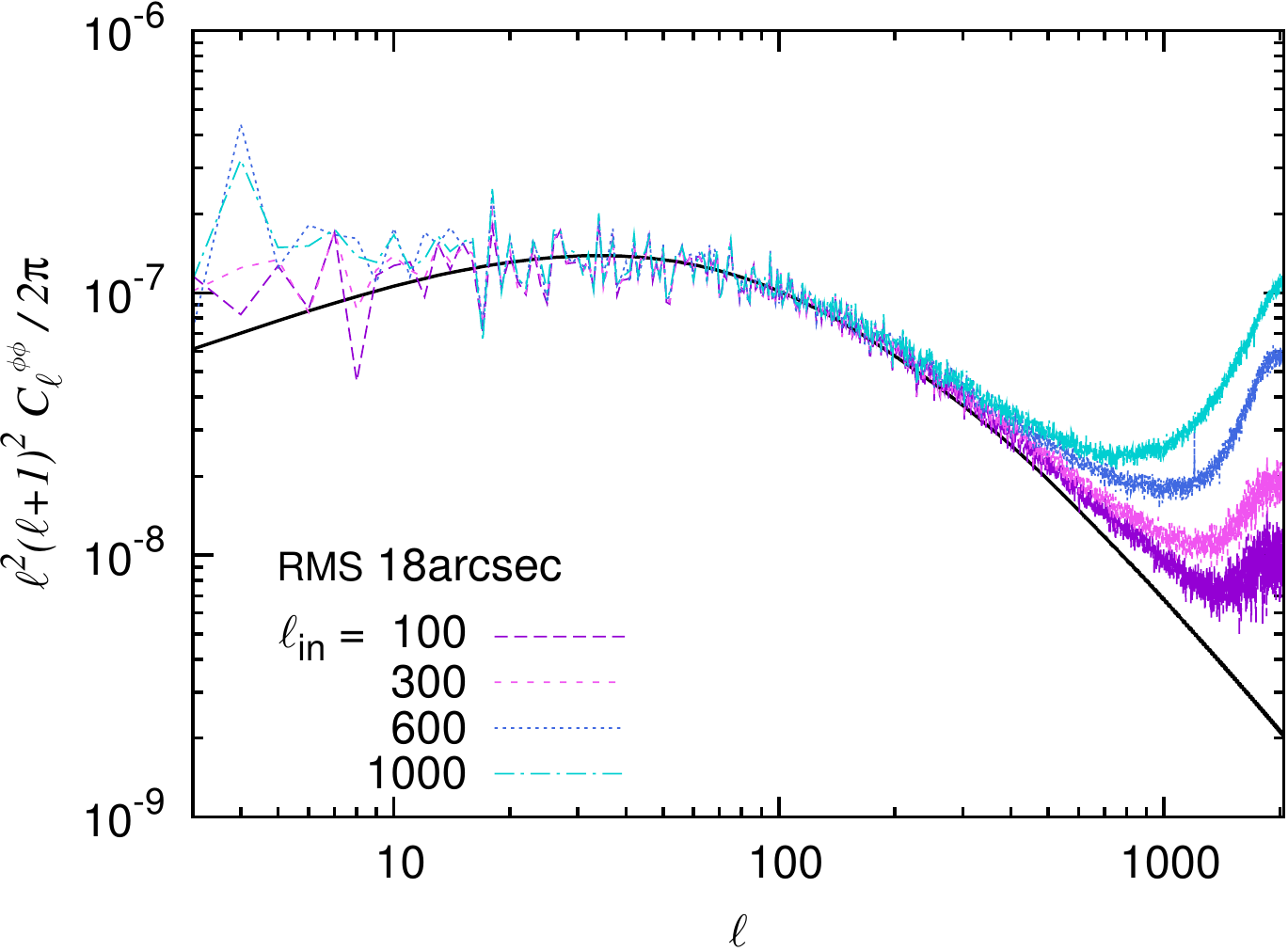}\label{clpp_rec_curl_s} }
\caption{Dependence on modulation scale ($\ell_{\rm in}$) of the curl-type pointing error. The thick black curve is the theoretical spectrum of the lensing $B$-modes. {\it Left}: angular power spectra of $B$-modes induced by the curl-type pointing error. {\it Right}: angular power spectra of lensing potentials reconstructed in presence of the curl-type pointing error. The thick black curve is the theoretical spectrum of the lensing potential.}
\label{curl_s}
\end{figure*}

The figure \ref{clbb_sys_grad_a} shows the angular power spectra of $B$-modes induced by the gradient-type pointing error with $\ell_{\rm in} = 600$. The RMS values of the error fields are $4$\,arcsec, $12$\,arcsec, and $36$\,arcsec. The figure \ref{clpp_rec_grad_a} shows the angular power spectra of the reconstructed lensing potentials in the respective cases. The mechanism of $E$ to $B$ leakage due to the gradient-type pointing error is mathematically equivalent to that of the gravitational lensing effects, and in fact the shapes of the induced $B$-mode spectra quite resemble that of the lensing $B$-mode spectrum. In the case of RMS 36arcsec, the induced $B$-mode spectrum is almost degenerated with the lensing B-mode spectrum. While 36arcsec is substantially smaller than 2arcmin, which is the RMS amplitude of deflection angles due to the gravitational lensing effects, the power of the pointing error is localized at $\ell_{\rm in}=600$ in this case (see also the next paragraph). The spectra of the reconstructed lensing potentials exhibit biases similar to those in the case of the gain error, i.e. spikes at multipoles relevant to imposed error fields and coherent deviation from the theoretical spectrum in the scale ranges of several degrees and sub-degree. These biases are not mere contamination in the lensing reconstruction analysis. Fields of the pointing error are simultaneously reconstructed and the biases in the reconstructed lensing potentials can be utilized for removing the pointing error from the polarization fields. In Sec. \ref{del_pointing}, we discuss a consequence of this property in the delensing analysis. 

The figure \ref{clbb_sys_grad_s} shows the angular power spectra of $B$-modes induced by the gradient-type pointing error, the fields of which have their RMS values fixed at $36$\,arcsec. The adopted multipoles of the error fields are $100$, $300$, $600$, and $1000$. The figure \ref{clpp_rec_grad_s} shows the angular power spectra of the reconstructed lensing potentials in the respective cases. Error fields with small  $\ell_{\rm in}$ cause spatially coherent displacement of the local structure in the CMB polarization field. When we further decrease $\ell_{\rm in}$ from $100$, the acoustic peaks in the CMB $E$-mode spectrum gradually appear in the induced $B$-mode spectrum. On the other hand, the induced B-modes grow as we increase $\ell_{\rm in}$ and the high-$\ell$ bias in the reconstructed lensing potentials increases accordingly. In addition, we can clearly see the correspondence between $\ell_{\rm in}$ and the multipoles of the spikes in Fig \ref{clpp_rec_grad_s}.

The figure \ref{clbb_sys_curl_a} shows the angular power spectra of $B$-modes induced by the curl-type pointing error with $\ell_{\rm in} = 600$. The RMS values of the error fields are $2$\,arcsec, $6$\,arcsec, and $18$\,arcsec. The figure \ref{clpp_rec_curl_a} shows the angular power spectra of the reconstructed lensing potentials in the respective cases. Unlike other error species, the induced $B$-mode power of this kind of error is localized in small scales for geometrical reason \cite{Namikawa2014b}. An induced $B$-mode spectrum for a given set of error parameters (i.e. an RMS value and $\ell_{\rm in}$) has a higher amplitude around $\ell=1000$ than that in the case of the gradient-type pointing error for the same parameter set, which results in more impacts on the subsequent lensing analysis. The spectra of the reconstructed lensing potentials are responding to the induced $B$-mode amplitude around $\ell=1000$ similarly to the case of the gain error though the spikes in the spectra are seen only at the overtone scale  as those shown in Fig. \ref{clpp_rec_angl_a} for the angle error (see also Fig. \ref{clpp_rec_curl_mf}).

The figure \ref{clbb_sys_curl_s} shows the angular power spectra of $B$-modes induced by the curl-type pointing error the fields of which have their RMS values fixed at $18$\,arcsec. The adopted multipoles of the error fields are $100$, $300$, $600$, and $1000$. The figure \ref{clpp_rec_grad_s} shows the angular power spectra of the reconstructed lensing potentials in the respective cases. The dependence on $\ell_{\rm in}$ is almost the same as that in the case of the gradient-type pointing error except for the low-$\ell$ behavior of the induced $B$-modes. 

\subsection{Mean-field bias} \label{rec_mf}

\begin{figure*}
\hspace{-13pt}
\subfloat[]{
\includegraphics[width=3.05in]{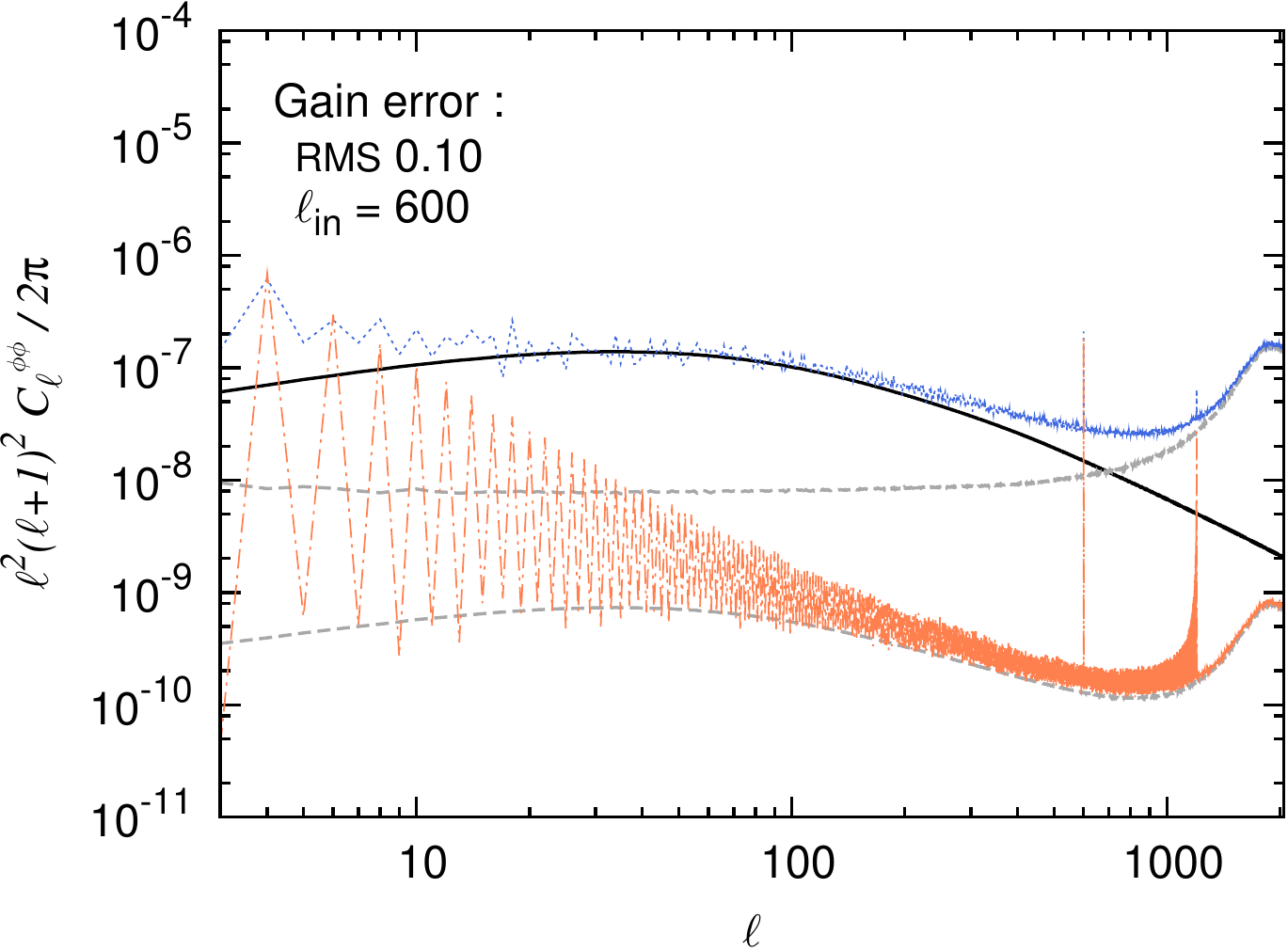}\label{clpp_rec_gain_mf} }
\hspace{-4pt}
\subfloat[]{
\includegraphics[width=3.05in]{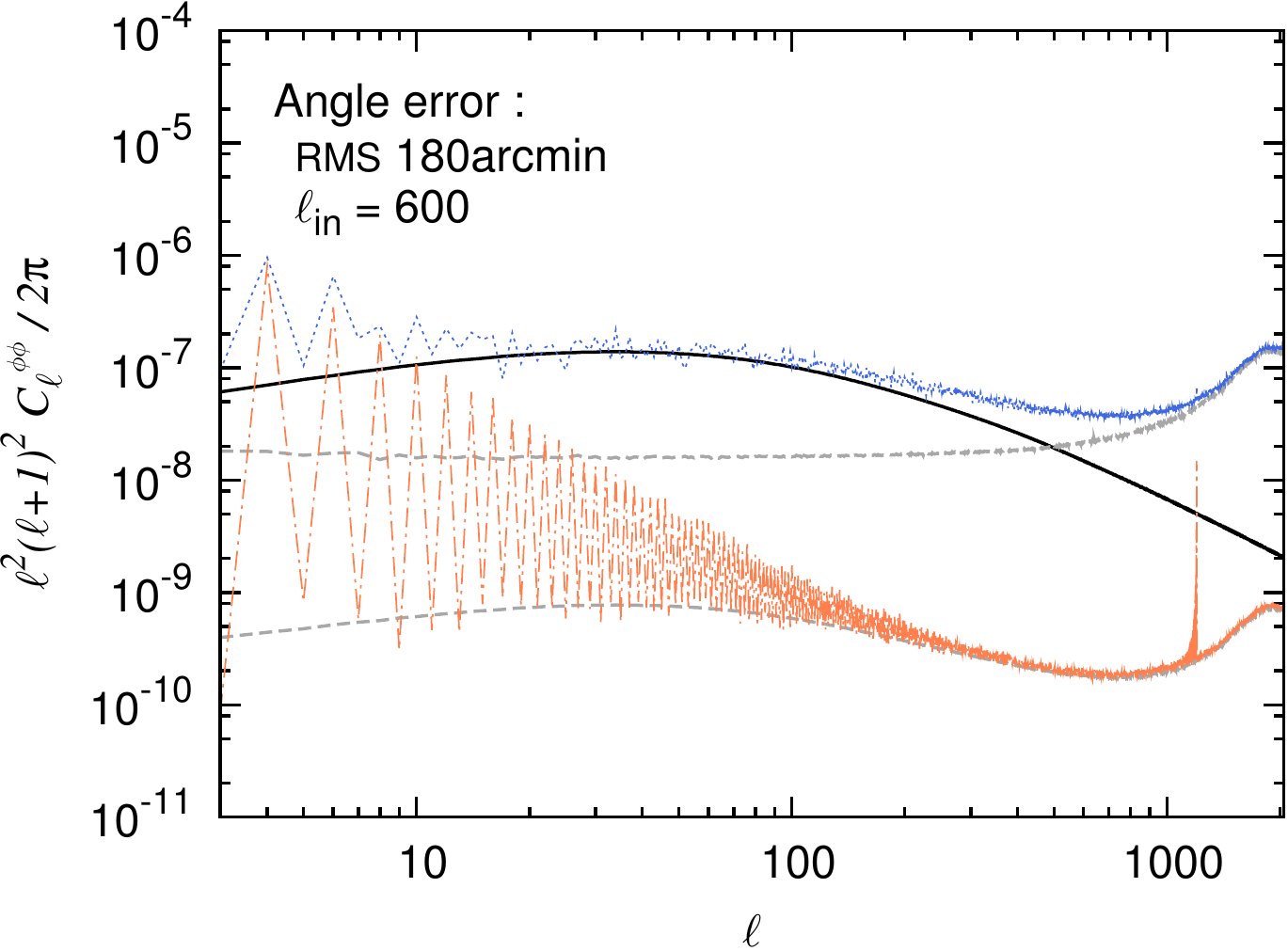}\label{clpp_rec_angl_mf} }

\hspace{-13pt}
\subfloat[]{
\includegraphics[width=3.05in]{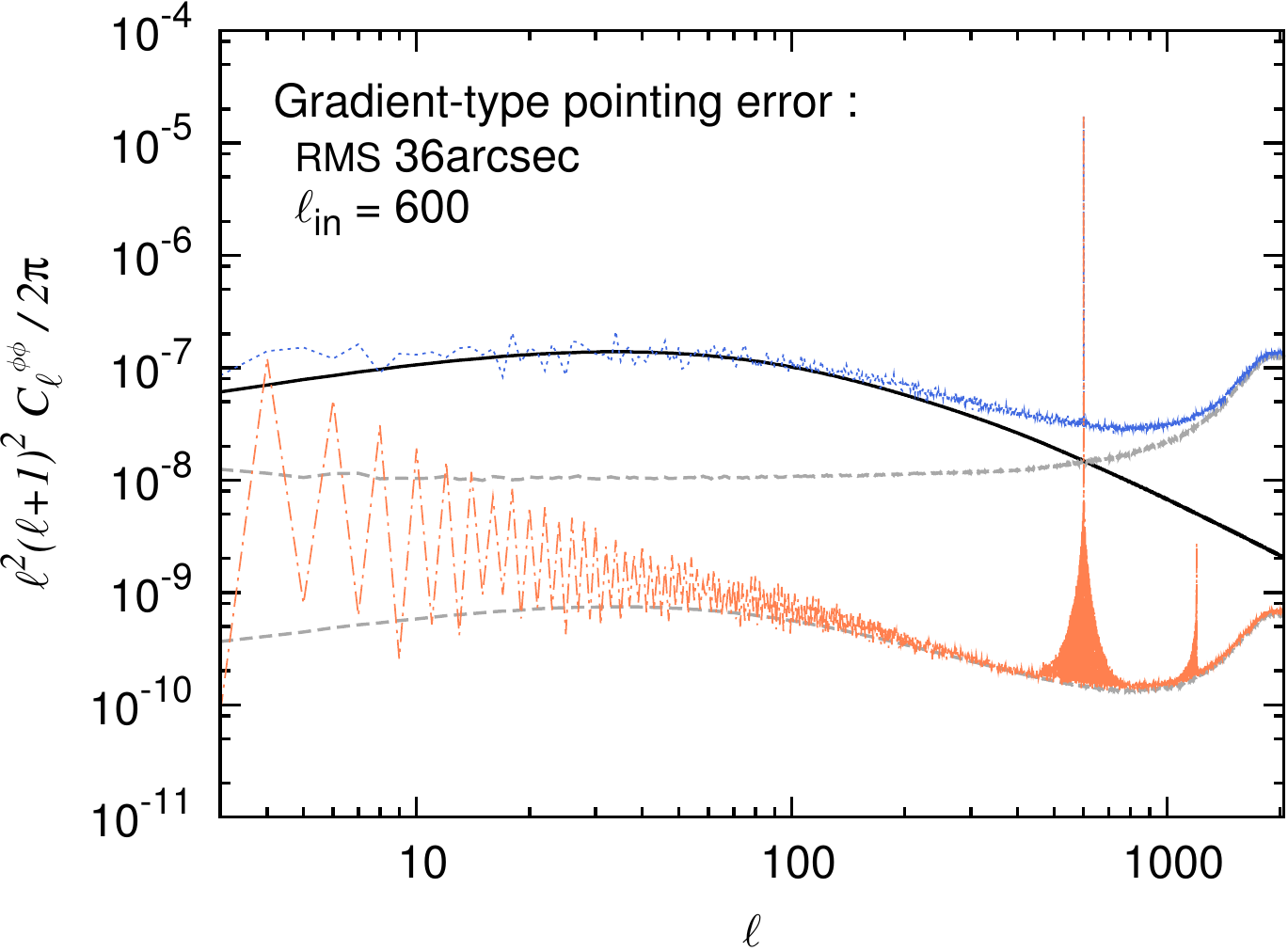}\label{clpp_rec_grad_mf} }
\hspace{-4pt}
\subfloat[]{
\includegraphics[width=3.05in]{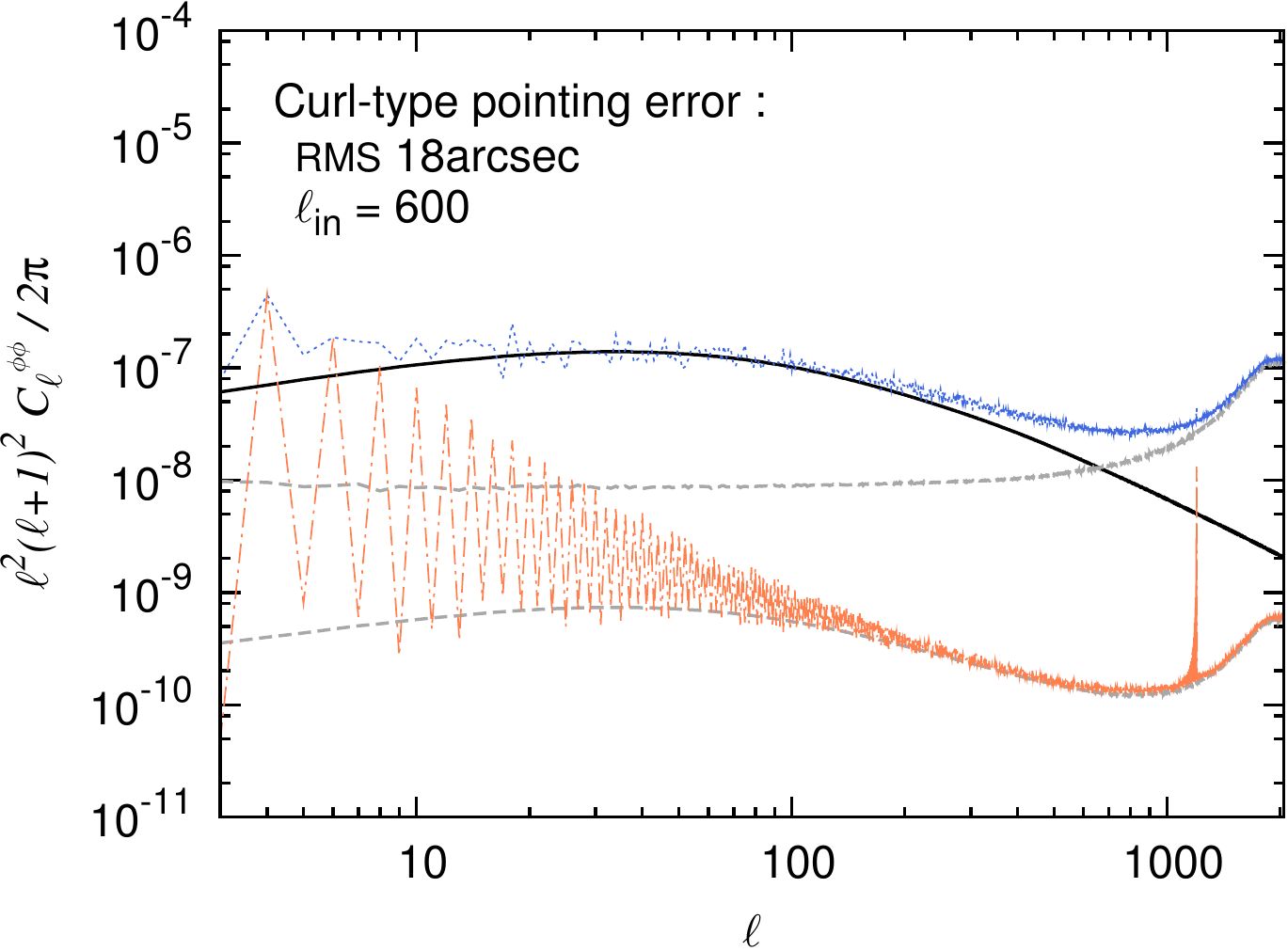}\label{clpp_rec_curl_mf} }
\caption{Spectra of mean-fields from the systematic errors (dot-dashed coral) compared with those of their respective reconstructed lensing potentials (dotted blue). The associated Gaussian biases are not subtracted from the spectra of the reconstructed lensing potentials. The thick black curve is the theoretical spectrum of the lensing potential. The upper and lower dashed gray curves are the Gaussian biases and spectra of Monte Carlo errors in the evaluated mean-fields, respectively.}
\label{clpp_rec_mf}
\end{figure*}

It is known that spatial modulations applied to the CMB polarization field make biases in reconstructed lensing potentials which are called ``mean fields''. In such a case, the estimator defined in Eq. \ref{EB-est} is a biased estimator and the associated angular power spectrum is biased by the contribution from a mean field. Some of the behavior observed in the spectra of the reconstructed lensing potentials shown in the previous subsections can be understood in terms of the mean-field bias. 

Formulation of the mean-field theory is found in literature, e.g. \cite{Namikawa:2012:bhe}. Unlike the simulations in the previous subsections, in the case of each error species we repeat the simulation procedure 200 times to obtain two sets of 100 Monte Carlo samples in which realizations of the CMB polarization are independent and parameters for the systematic error are fixed. The mean field of the reconstructed lensing potential is evaluated as the average of the 200 samples of the estimator. Also, we obtain the associated Gaussian bias by cross-correlating $E$-modes in the one sample set with $B$-modes in the other sample set or vice versa. The figures \ref{clpp_rec_gain_mf} to \ref{clpp_rec_curl_mf} show the spectra of the mean fields from the systematic errors discussed in the previous subsections. 

It is clearly found that the biases in the large scales and the spikes at the multipoles relevant to the imposed error fields are manifestations of the mean fields. While details of the mean fields exhibit some complexity and depend also on spatial configurations of the error fields, these features are robust and common to them. On the other hand, we can see that the mean fields do not cause the coherent deviations in the small scales which have major impacts on the delensing analysis. Instead, they are understood to be due to the biases in high-$\ell$ power in the respective $B$-mode spectra which increase the reconstruction noise  relevant to the Gaussian biases. More specifically, they do not come from biases in the reconstructed lensing potentials but from increase in variance around the true lensing potential. In addition, by performing mean-field subtraction from the reconstructed lensing potentials, we confirmed that the spikes also contribute somewhat to degradation in delensing efficiency depending on their amplitude and multipoles. Note that the subtraction is not actually applied in the delensing analysis described in the next section (see the paragraph below). 

In the Monte Carlo simulations, we fixed the configurations of the error fields to extract their respective mean fields. However, in practical situations, the systematic errors discussed here are not expected to make such fixed modulations predictable in advance of observation. Mitigation of mean fields from potential systematic errors is a considerable task to be addressed. The bias hardening \cite{Namikawa:2012:bhe,Namikawa:2013:bhepol} is a possible method to do it.

\section{Delensing} \label{delensing}
In this section, we compare results of the delensing analysis in the cases of the systematic errors described in the previous sections. Residual lensing $B$-modes after delensing are defined in Eq. \ref{B-residual} and we make the multipole moments up into the associated angular power spectrum ($\widehat{C}_{\ell}^{BB,{\rm res}}$).  We define the delensing efficiency as the fractional difference of this power spectrum from the theoretical spectrum of the lensing $B$-modes. 

\subsection{Gain error} \label{del_gain}

\begin{figure*}
\hspace{-13pt}
\subfloat[]{
\includegraphics[width=3.05in]{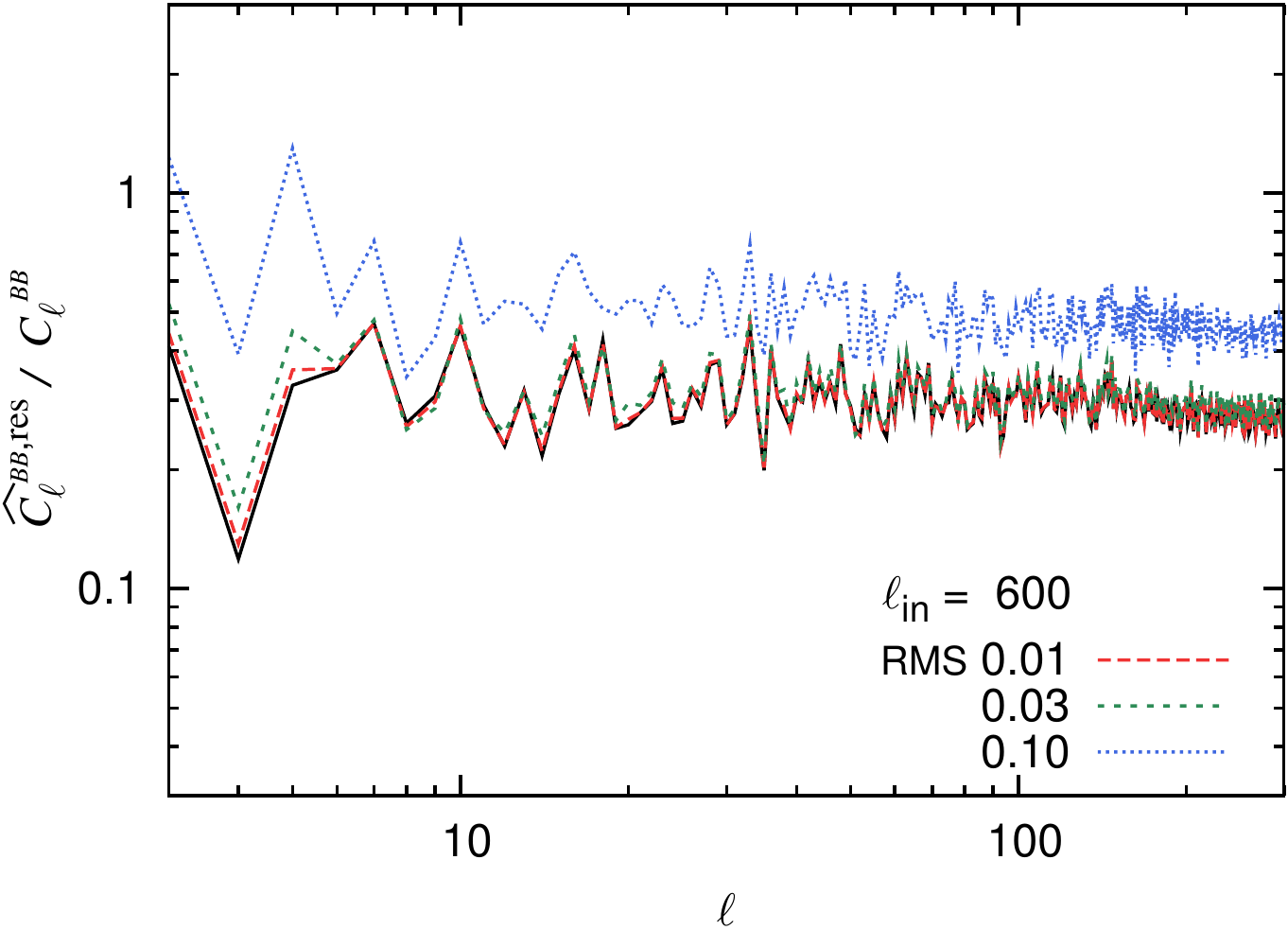}\label{clbb_del_gain_a} }
\hspace{-4pt}
\subfloat[]{
\includegraphics[width=3.05in]{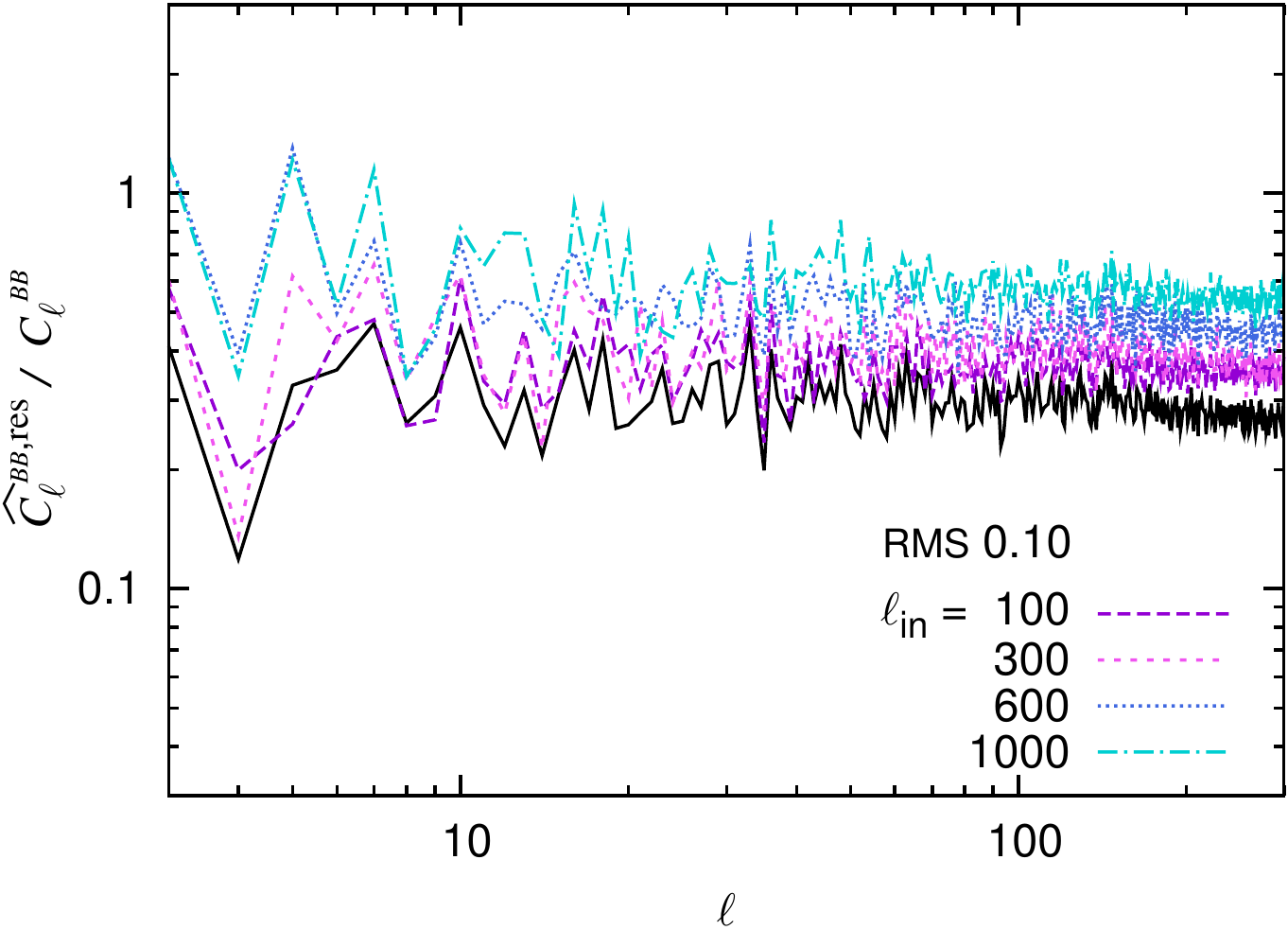}\label{clbb_del_gain_s} }
\caption{ Residual fractions of the lensing $B$-modes in the case of the gain error. The thick black line is that in the case without any systematic errors. {\it Left}: dependence on RMS amplitude of the gain error. {\it Right}: dependence on modulation scale of the gain error.}
\label{figs_del_gain}
\end{figure*}

The figure \ref{clbb_del_gain_a} shows the residual fractions of the lensing $B$-modes after delensing performed with lensing potentials reconstructed in the presence of the gain error. They correspond to the cases shown in the figure \ref{gain_a}. The figure \ref{clbb_del_gain_s} shows those which correspond to the cases in the figure \ref{gain_s}. The levels of the overall amplitude of the residual lensing $B$-modes totally depend on the bias levels of the respective reconstructed lensing potentials. The spectrum of the residual lensing $B$-modes in the case of RMS $0.01$ is indistinguishable from that in the case without any systematic errors. The amplitude of the spectrum is about $25\%$ of that of the original lensing $B$-modes. With an increase in the RMS value from $0.01$, the delensing efficiency begins to degrade around RMS $0.03$, in the case of which the amplitude of the induced $B$-mode spectrum is about $10\%$ of that of the lensing $B$-modes. 

It is noted that the delensing efficiency in the case of $0.10$ RMS and $\ell_{\rm in} = 600$ is substantially degraded but still about $50\%$. In that case, the residual lensing $B$-modes are much smaller than the induced $B$-modes, which are comparable to the lensing $B$-modes (see Fig. \ref{clbb_sys_gain_a}). In such a case, the importance of systematics removal is comparable to that of delensing. In addition to conventional error correction, there may be a possibility of reconstruction of the systematic errors which is formulated in the same way as the lensing reconstruction analysis \cite{Yadav2010}. However, it is not a trivial problem to find an appropriate filter function such as the Wiener filter in the case of the delensing analysis. An example of tackling the problem is found in Ref. \cite{Williams2021}.

\subsection{Angle error} \label{del_angle}

\begin{figure*}
\hspace{-13pt}
\subfloat[]{
\includegraphics[width=3.05in]{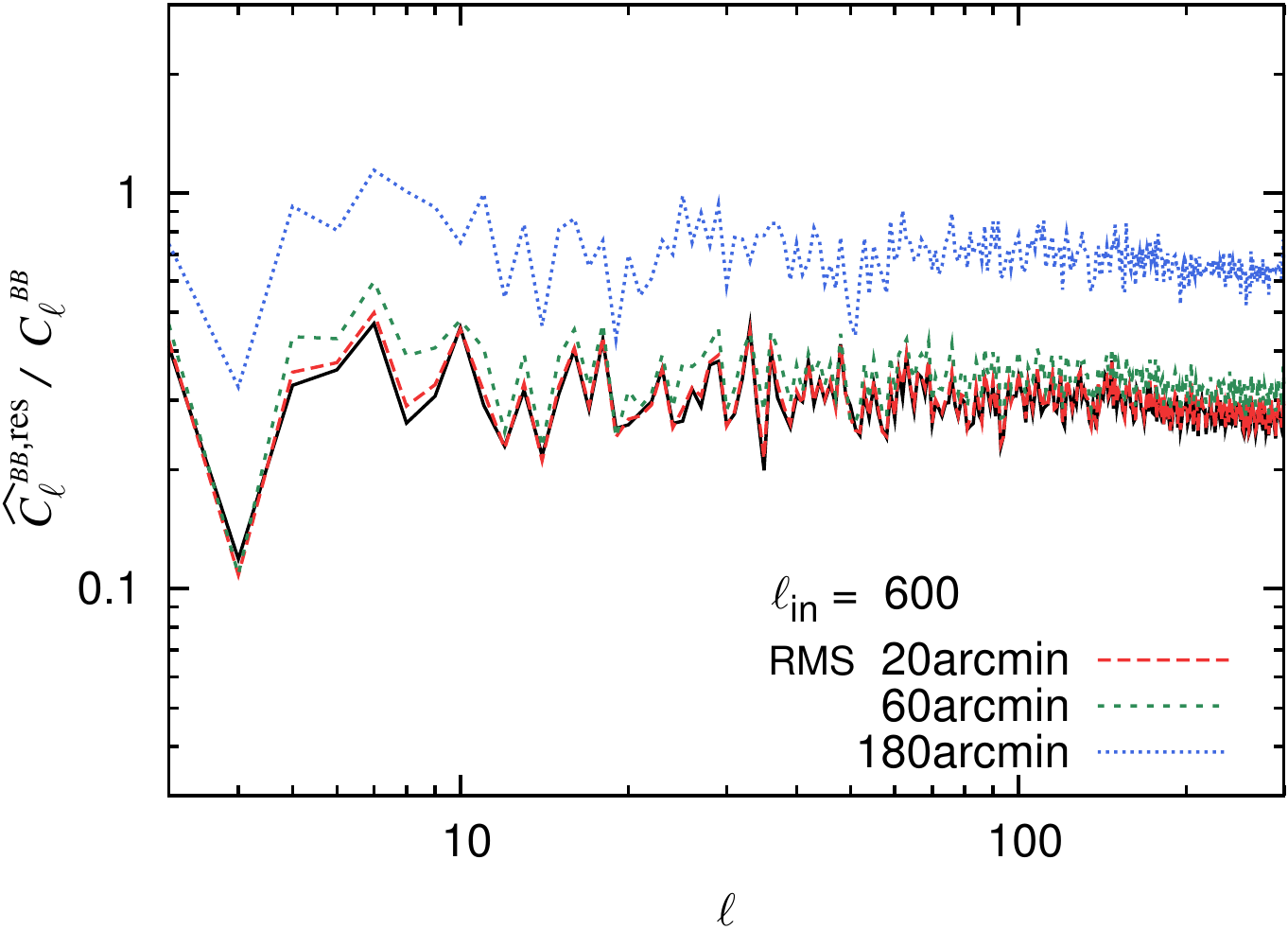}\label{clbb_del_angl_a} }
\hspace{-4pt}
\subfloat[]{
\includegraphics[width=3.05in]{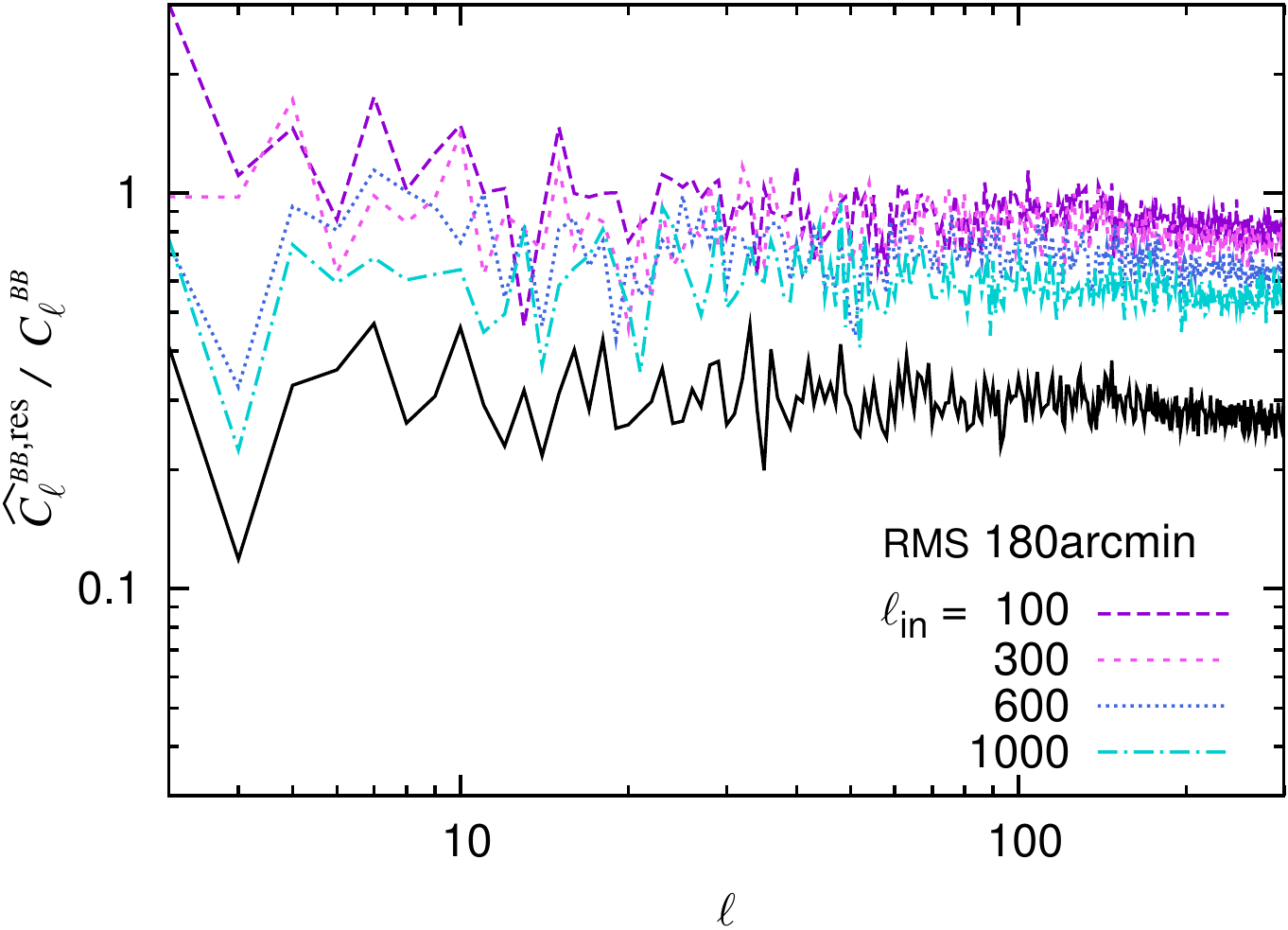}\label{clbb_del_angl_sh} }
\caption{ Residual fractions of the lensing $B$-modes in the case of the angle error. The thick black line is that in the case without any systematic errors. {\it Left}: dependence on RMS amplitude of the angle error. {\it Right}: dependence on modulation scale of the angle error. }
\label{del_angl}
\end{figure*}

The figure \ref{clbb_del_angl_a} shows the residual fractions of the lensing $B$-modes after delensing performed with lensing potentials reconstructed in the presence of the angle error. They correspond to the cases shown in the figure \ref{angl_a}. The figure \ref{clbb_del_angl_sh} shows those which correspond to the cases in the figure \ref{angl_sh}. The behavior of the residual lensing $B$-modes is similar to that in the case of the gain error though their dependence on $\ell_{\rm in}$ is in the opposite sense. These are implied by the figures \ref{clpp_rec_gain_a}, \ref{clpp_rec_gain_s}, \ref{clpp_rec_angl_a}, and \ref{clpp_rec_angl_sh}. 

In the case of RMS $180$arcmin and $\ell_{\rm in} = 600$, the residual lensing $B$-modes are definitely smaller than the induced $B$-modes (cf. Fig. \ref{clbb_sys_angl_a}). There are examples of angle error reconstruction applied to real observation data, which illustrates the effectiveness of such methods and the possibility of further removing false signals in the $B$-modes \cite{Minami2020,ACT2020}.

\subsection{Pointing error} \label{del_pointing}

\begin{figure*}
\hspace{-13pt}
\subfloat[]{
\includegraphics[width=3.05in]{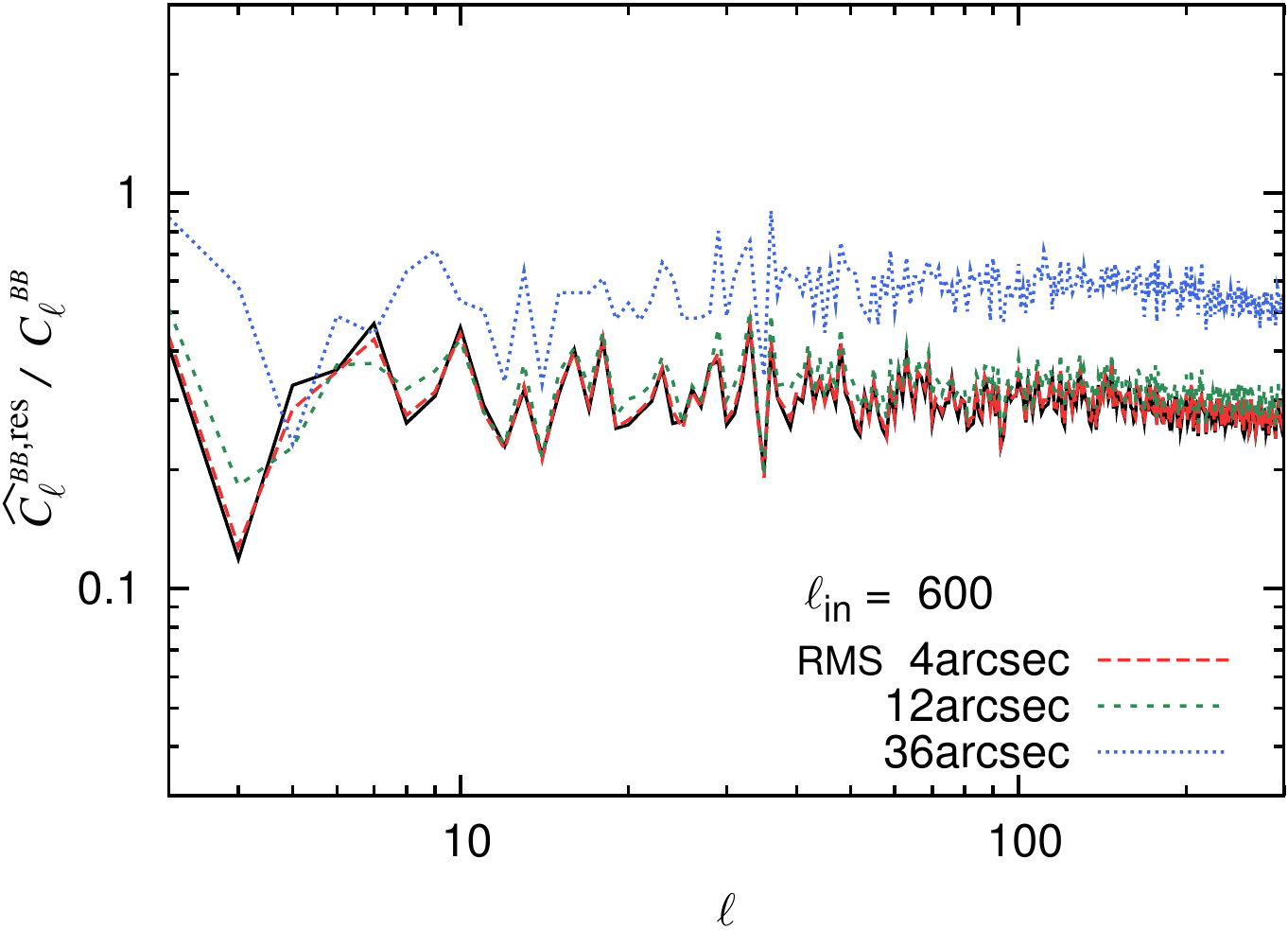}\label{clbb_del_grad_a} }
\hspace{-4pt}
\subfloat[]{
\includegraphics[width=3.05in]{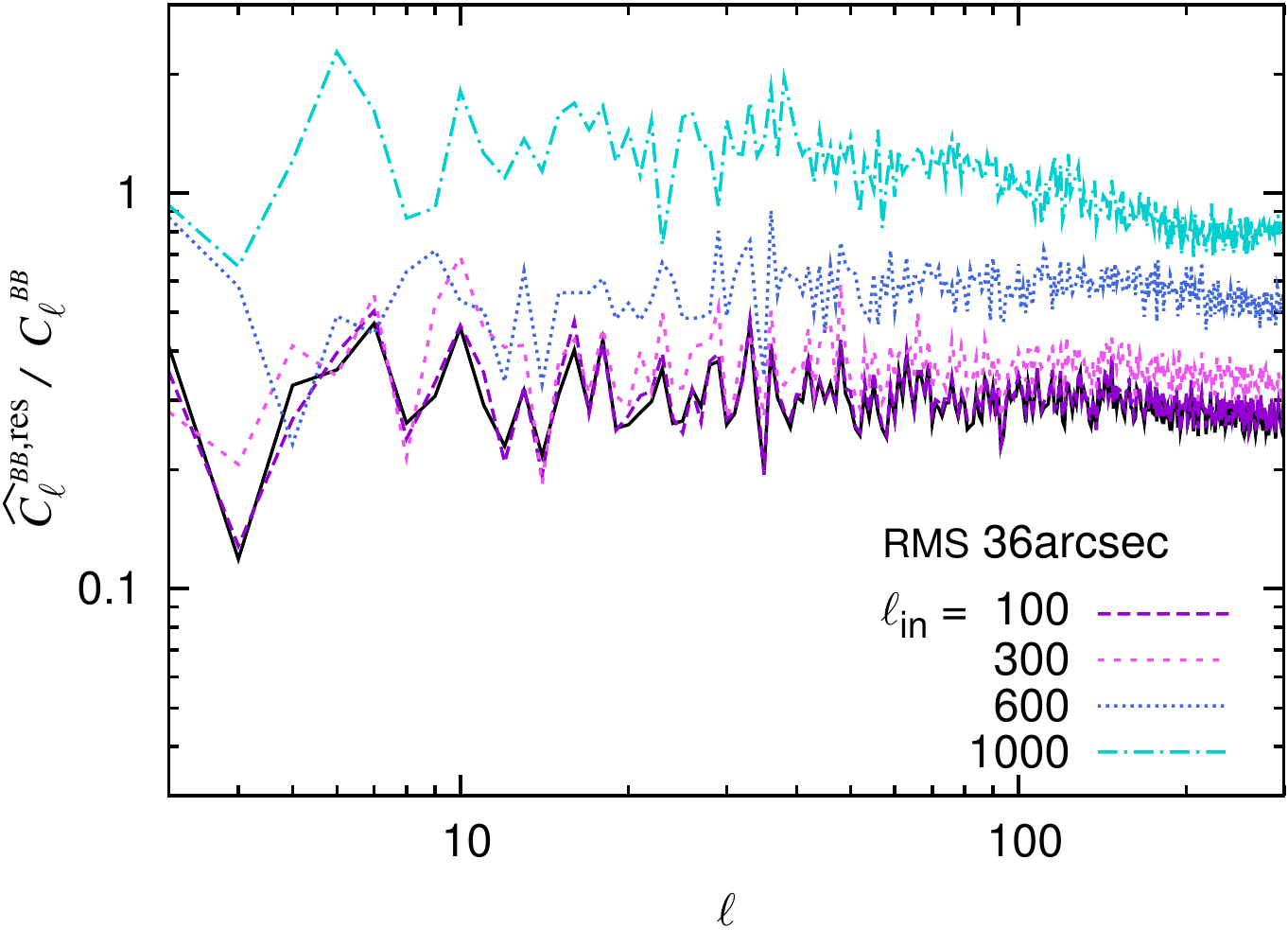}\label{clbb_del_grad_s} }
\caption{ Residual fractions of the lensing $B$-modes in the case of the gradient-type pointing error. The thick black line is that in the case without any systematic errors. {\it Left}: dependence on RMS amplitude of the gradient-type pointing error. {\it Right}: dependence on modulation scale of the gradient-type pointing error.}
\label{del_grad}
\end{figure*}

\begin{figure*}
\hspace{-13pt}
\subfloat[]{
\includegraphics[width=3.05in]{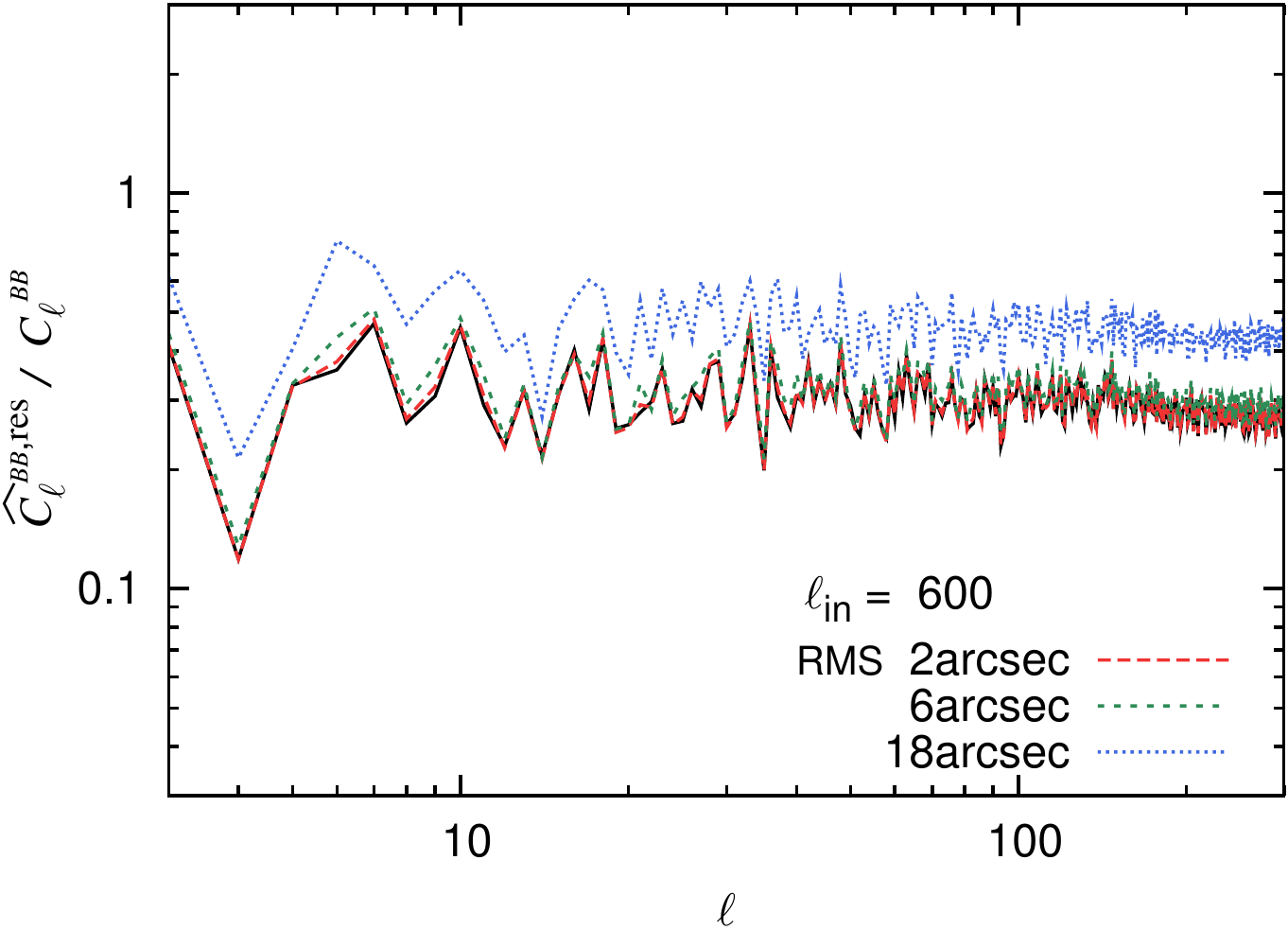}\label{clbb_del_curl_a} }
\hspace{-4pt}
\subfloat[]{
\includegraphics[width=3.05in]{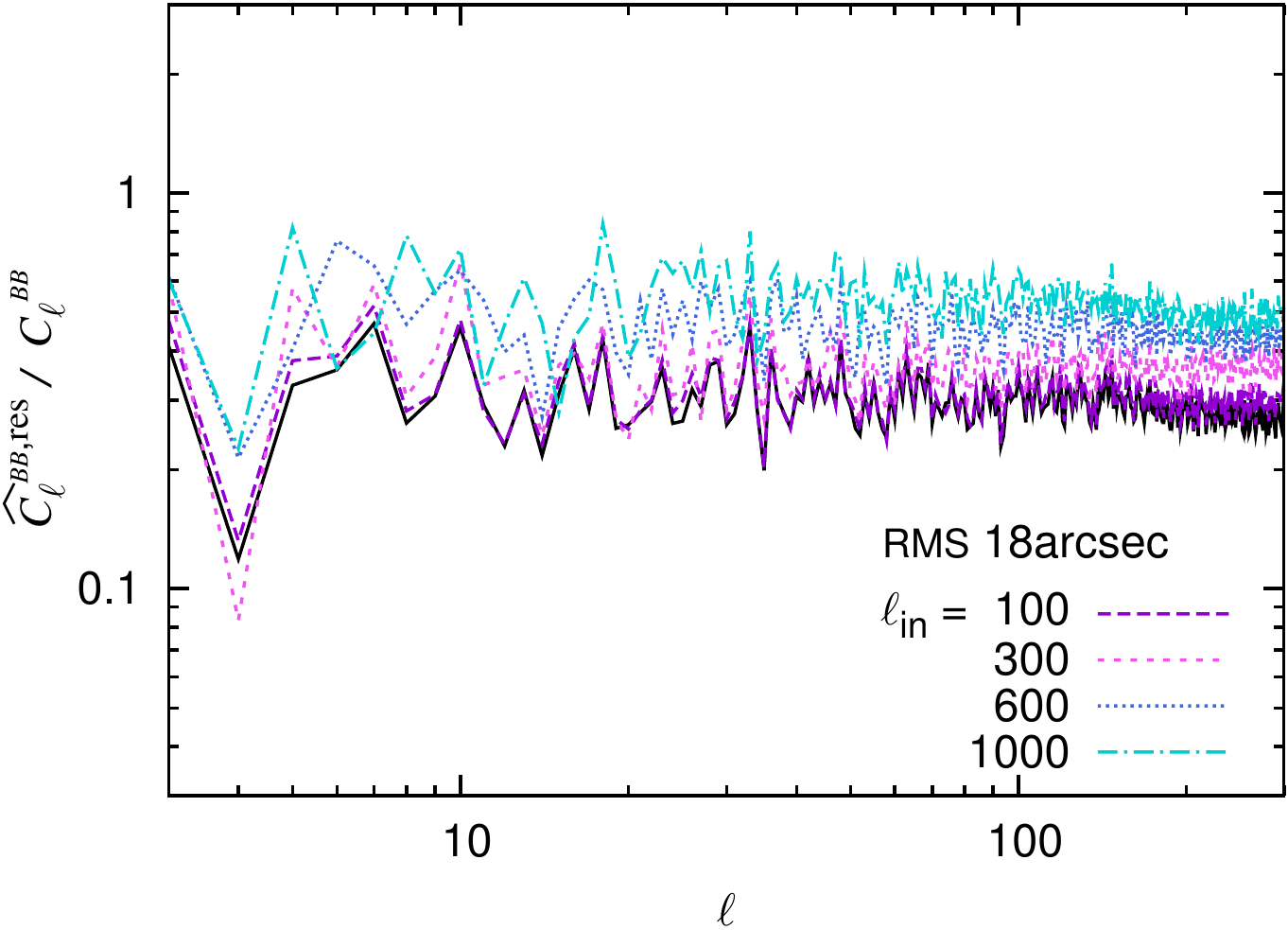}\label{clbb_del_curl_s} }
\caption{ Residual fractions of the lensing $B$-modes in the case of the curl-type pointing error. The thick black line is that in the case without any systematic errors. {\it Left}: dependence on RMS amplitude of the curl-type pointing error. {\it Right}: dependence on modulation scale of the curl-type pointing error.}
\label{del_curl}
\end{figure*}

The figure \ref{clbb_del_grad_a} shows the residual fractions of the lensing $B$-modes after delensing performed with lensing potentials reconstructed in the presence of the gradient-type pointing error. They correspond to the cases shown in the figure \ref{grad_a}. The figure \ref{clbb_del_grad_s} shows those which correspond to the cases in the figure \ref{grad_s}. While the residual lensing $B$-modes mostly show similar behavior to that in the case of the gain error, the dependence on $\ell_{\rm in}$ is slightly stronger. This is related to bias in the spectral amplitude of the reconstructed lensing potentials at multipoles of several hundreds, which can be seen in Figs. \ref{clpp_rec_gain_s} and \ref{clpp_rec_grad_s}. The presence of the higher spikes also contributes to such scale dependence (see Fig. \ref{clpp_rec_mf}). Unlike other error species, the gradient-type pointing error has the same properties as the gravitational lensing effects. In the lensing reconstruction analysis, the error fields are also reconstructed simultaneously with the lensing potentials. Since the estimators contain the contributions from the pointing error in the proper way, we subtract the template $B$-modes from the $B$-modes including the pointing error contributions instead of the true lensing $B$-modes themselves. Without such treatment, we suffer over-subtraction which causes bias in the spectra of the residual lensing $B$-modes. This means that the usual lensing analysis which is not specialized for error correction has the power to reject this kind of pointing error (partially, at least).

The figure \ref{clbb_del_curl_a} shows the residual fractions of the lensing $B$-modes after delensing performed with lensing potentials reconstructed in the presence of the curl-type pointing error. They correspond to the cases shown in the figure \ref{curl_a}. The figure \ref{clbb_del_curl_s} shows those which correspond to the cases in the figure \ref{curl_s}. In this case, the responses of the residual $B$-modes to the imposed error fields are also similar to those in the case of the gain error. Although the induced $B$-modes lack their power in large scales as shown in Figs. \ref{clbb_sys_curl_a} and \ref{clbb_sys_curl_s}, we just mention the fact that the reconstruction analysis of curl-type gravitational lensing effects, which is formulated for the purpose of detecting the primordial GWs or other objects of exotic origin \cite{Cooray2005,Namikawa2012}, can be applied for removing this systematic error.

\section{Summary and discussion} \label{summary}
We have illustrated the effects of the observational systematic errors on the lensing analysis of the CMB polarization. We considered three kinds of systematic errors, i.e. gain error, angle error, and pointing error, which were modeled in terms of error fields. A suite of simulations, which consists of generating a modulated polarization field, lensing reconstruction, and subsequent delensing, was performed for each error species, and this was repeated with different values of the error parameters RMS amplitude and a multipole of a single non-vanishing multipole moment of the error field.

The systematic errors cause biases in reconstructed lensing potentials. Spatial modulations due to the systematic errors result in mean fields. The mean fields exhibit a sharp response to multipole moments of imposed error fields at multipoles relevant to the multipole moments. The mean fields also appear as biases in reconstructed lensing potentials in large scales. Apart from mean fields, additional variance in reconstructed lensing potentials in small scales comes from extra $B$-mode power induced by the systematic errors, and this ends up being a major contaminant for the delensing analysis. 

Impacts on the results of the lensing analysis depend on the amplitude of the systematic errors. If the amplitude of the induced $B$-modes is smaller than a few percent of that of the lensing $B$-modes in the power spectrum domain, there is little impact on the lensing analysis. As described in Sec. \ref{method}, the error fields discussed in this work stand for the net errors at each sky position, which are the averages of instantaneous errors. Given that fact, some of the RMS error values adopted in the sections \ref{reconstruction} and \ref{delensing} are excessively large. In actual observations, the levels of gain and angle accuracy have already reached sub-percent and a few dozen arcmin, respectively \cite{PARAde2014,Louis2017,PARAde2017}. Also, planned future projects are targeting a pointing accuracy of a few arcsec \cite{PICO2019,S42019}. However, in actual data analysis, we should be aware that there may be possibilities of other indirect influence such as the derivation of errors of different species from the errors concerned.

In practical situations, the power of the error fields does not localize in a narrow multipole range and their spectra are usually multi-modal. In such cases, the whole power of the error fields is divided into small pieces and these are distributed to many multipole moments each of which makes a small contribution to the induced $B$-modes. Besides this, multipole moments to which the induced $B$-modes do not sensitively respond also have finite power. For example, in the case of the EPIC baseline scan \cite{EPIC2008}, a pointing error which is assumed to be static in the telescope frame results in an error field dominated by the curl-type pointing error and the spectrum consists of a white noise component, a scan synchronous component associated with its harmonics, and an inverse power-law component which is dominant in large scales. With a normalization of RMS 18arcsec applied to the error field, the spectrum of the induced $B$-modes has amplitude smaller than that of the lensing $B$-modes by about one order (cf. Fig. \ref{clbb_sys_curl_a} and \ref{clbb_sys_curl_s}). Meanwhile, some trends of the errors concerned in this paper are found in past observations. Gain fluctuations, which are often discussed in connection with $1/f$ noise, are usually modeled and identified as coherent variations over observation areas \cite{Seiffert2002,Buder2012,Planck2013LFIcal}. After gain calibration, residual gain errors have their power biased to scales comparable to observation areas (otherwise calibration intervals) and likely make red-tilted components in polarization spectra\cite{PARAde2014,PARAde2017,PlanckEarlyLFI2011,QUIET2011,Planck2015LFIproc,Planck2018HFIproc,PB2020}. Angle errors tend to be evaluated as being static and therefore homogeneous in most cases because inhomogeneous angle errors have less impact on polarization fields. In addition, there are few reports indicating that short-time angle jitters have a notable effect \cite{ACT2020,Buder2012,PB2020}. While large parts of the whole power which make significant contributions to polarization fields are expected to reside in the largest scales of observation areas, they have been mostly removed by angle correction techniques \cite{Millea:2020:MLEphi,PARAde2014,PARAde2017,PB2020,Kaufman2014,Naess2014,Choi2020}. Pointing errors are generally argued to have both long-time coherence and almost random components \cite{Mirmelstein2020,Planck2015LFIsys}. The former which come from, for example, mis-calibration of pointing models, can have complicated spectra, as discussed above in the case of EPIC, and their amplitude would be mitigated by carefully designed scan patterns. Several reports indicate that such components tend to form significant parts \cite{QUIET2011,PB2020,Planck2015LFIsys,Planck2013HFIproc,Crites2015}. The latter raise levels of white noise components although they are severely suppressed due to the averaging-out effect of multiple sampling.

Finally, we comment on cases with observation noise. The inclusion of observation noise into our simulations leads to an increase in the variance of reconstructed lensing potentials. Unlike the systematics-induced biases, statistical properties of the additional variance are theoretically predictable on the basis of experimental specifications. Due to the additional variance, fractional contributions from the systematic errors to whole reconstruction errors are reduced. Consequently, the difference in delensing efficiency between cases with and without the systematic errors decreases in the presence of observation noise.

\section*{Acknowledgments}

We acknowledge the use of {\tt CAMB} \cite{Lewis:1999bs}, {\tt Healpix} \cite{Gorski:2004by}, and {\tt Lenspix} \cite{Challinor:2005jy}. This work is supported by JSPS KAKENHI Grant Number JP17K05478.

\bibliography{refs}

\newpage

\appendix

\section{Induced $E$- and $B$-modes} \label{eqs-EBlm}
For readers' convenience, we show mathematical expressions of $E$- and $B$-modes induced by the systematic errors described in Sec. \ref{method_error}. By denoting the difference between a modulated polarization field and its original field as $\delta [Q \pm iU](\hat{\mbox{\boldmath $n$}})$, we define the induced $E$- and $B$-modes as
\begin{equation}\label{EMlm-def}
\begin{split}
\delta [ E \pm i B ]_{\ell m} = - \int d\hat{\mbox{\boldmath $n$}} \, _{\pm2} Y^*_{\ell m}(\hat{\mbox{\boldmath $n$}}) \, \delta [Q \pm iU](\hat{\mbox{\boldmath $n$}}) .
\end{split}
\end{equation}
Those for the gain error, angle error, gradient-type pointing error, and curl-type pointing error are shown in Eq. \ref{EBlm-gain} to Eq. \ref{EBlm-curl} \cite{Hu2003,Namikawa2012,mccallum2020,zhai2020}. 
\begin{eqnarray}
\delta [ E \pm i B ]_{\ell m} = \sum_{\ell' m'} \sum_{\ell'' m''} g_{\ell' m'} [ E \pm i B ]_{\ell'' m''} \sqrt{\frac{(2\ell+1)(2\ell'+1)(2\ell''+1)}{4\pi}} \hspace{100pt} \nonumber \\
 \times (-1)^m  
  \left(
   \begin{array}{ccc}
    \ell & \ell' & \ell'' \\
    \pm 2 & 0 & \mp 2 \\
   \end{array}
  \right) 
  \left(
   \begin{array}{ccc}
    \ell & \ell' & \ell'' \\
    -m & m' & m'' \\
   \end{array}
  \right) . \hspace{10pt} \label{EBlm-gain} \\
\nonumber \\
\nonumber \\
\delta [ E \pm i B ]_{\ell m} = \pm 2i \sum_{\ell' m'} \sum_{\ell'' m''} \alpha_{\ell' m'} [ E \pm i B ]_{\ell'' m''} \sqrt{\frac{(2\ell+1)(2\ell'+1)(2\ell''+1)}{4\pi}} \hspace{78pt} \nonumber \\
 \times (-1)^m  
  \left(
   \begin{array}{ccc}
    \ell & \ell' & \ell'' \\
    \pm 2 & 0 & \mp 2 \\
   \end{array}
  \right) 
  \left(
   \begin{array}{ccc}
    \ell & \ell' & \ell'' \\
    -m & m' & m'' \\
   \end{array}
  \right)
+ \mathcal{O}(\alpha^2) . \hspace{10pt} \label{EBlm-angle} \\
\nonumber \\
\nonumber \\
\delta [ E \pm i B ]_{\ell m} = \frac{1}{2} \sum_{\ell' m'} \sum_{\ell'' m''} \psi_{\ell' m'} [ E \pm i B ]_{\ell'' m''}  \sqrt{\frac{\ell'(\ell'+1)(2\ell+1)(2\ell'+1)(2\ell''+1)}{4\pi}} \hspace{46pt} \nonumber \\
 \times (-1)^{m+1} \Biggl[
 \sqrt{(\ell''+2)(\ell''-1)}
  \left(
   \begin{array}{ccc}
    \ell & \ell' & \ell'' \\
    \pm 2 &  \mp 1 &  \mp 1 \\
   \end{array}
  \right) 
  \left(
   \begin{array}{ccc}
    \ell & \ell' & \ell'' \\
    -m & m' & m'' \\
   \end{array}
  \right)
  \hspace{62pt} \nonumber \\
 + \sqrt{(\ell''+3)(\ell''-2)}
  \left(
   \begin{array}{ccc}
    \ell & \ell' & \ell'' \\
    \pm 2 &  \pm 1 &  \mp 3 \\
   \end{array}
  \right) 
  \left(
   \begin{array}{ccc}
    \ell & \ell' & \ell'' \\
    -m & m' & m'' \\
   \end{array}
  \right) 
 \Biggr] + \mathcal{O}(\psi^2) . \hspace{10pt} \label{EBlm-grad} \\
\nonumber \\
\nonumber \\
\delta [ E \pm i B ]_{\ell m} = \pm \frac{i}{2} \sum_{\ell' m'} \sum_{\ell'' m''} \varpi_{\ell' m'} [ E \pm i B ]_{\ell'' m''}  \sqrt{\frac{\ell'(\ell'+1)(2\ell+1)(2\ell'+1)(2\ell''+1)}{4\pi}} \hspace{36pt} \nonumber \\
 \times (-1)^{m+1} \Biggl[ 
\sqrt{(\ell''+2)(\ell''-1)}
  \left(
   \begin{array}{ccc}
    \ell & \ell' & \ell'' \\
    \pm 2 &  \mp 1 &  \mp 1 \\
   \end{array}
  \right) 
  \left(
   \begin{array}{ccc}
    \ell & \ell' & \ell'' \\
    -m & m' & m'' \\
   \end{array}
  \right) \hspace{62pt} \nonumber \\
 - \sqrt{(\ell''+3)(\ell''-2)}
  \left(
   \begin{array}{ccc}
    \ell & \ell' & \ell'' \\
    \pm 2 &  \pm 1 &  \mp 3 \\
   \end{array}
  \right) 
  \left(
   \begin{array}{ccc}
    \ell & \ell' & \ell'' \\
    -m & m' & m'' \\
   \end{array}
  \right)
 \Biggr] + \mathcal{O}(\varpi^2) . \hspace{10pt} \label{EBlm-curl}
\end{eqnarray}
$g_{\ell m}$, $\alpha_{\ell m}$, $\psi_{\ell m}$, and $\varpi_{\ell m}$ are harmonic coefficients of $g(\hat{\mbox{\boldmath $n$}})$, $\alpha(\hat{\mbox{\boldmath $n$}})$, $\psi(\hat{\mbox{\boldmath $n$}})$, and $\varpi(\hat{\mbox{\boldmath $n$}})$, respectively. Except for the case of the gain error, these expressions have nonlinear terms. In the actual simulations, systematics imposition is performed in the map domain and the nonlinearities are taken into account. 

\end{document}